\documentclass[aps,reprint,amsmath,amssymb,superscriptaddress]{revtex4-2}
\usepackage{CJK}
\usepackage{dcolumn}
\usepackage{bm}
\usepackage{xspace}
\usepackage{amsmath}
\usepackage{graphicx}
\usepackage[caption=false]{subfig}
\usepackage{float}

\newcommand{\braket}[2] {{\langle #1 | #2 \rangle}}
\newcommand{\xh}{{\hat{x}}}

\newcommand{\zh}{{\hat{z}}}
\newcommand{\nh}{{\hat{n}}}
\newcommand{\mv}{{\bf m}}

\newcommand{\bdelta}{{\boldsymbol \delta}}

\newcommand{\kv}{{\bf k}}
\newcommand{\Rv}{{\bf R}}

\newcommand{\Hv}{{\bf H}}
\newcommand{\Sv}{{\bf S}}
\newcommand{\Mv}{{\bf M}}
\newcommand{\Keff}{{K_{\rm eff}}}
\newcommand{\Kexp}{{K_{\rm exp}}}
\newcommand{\crte}{1T-CrTe$_2$\xspace}
\newcommand{\ajd}{{a_j^\dagger}}

\usepackage[unicode=true,
 bookmarks=false,
 breaklinks=false,pdfborder={0 0 1},colorlinks=true] 
 {hyperref}
\hypersetup{pdfstartview=FitV,linkcolor=blue,citecolor=blue,urlcolor=blue}

\begin{document}
\begin{CJK*}{GB}{}

\title{Structural, electronic, and magnetic properties of CrTe$_2$}

\author{Yuhang Liu}
	\email{yliu446@ucr.edu}
	\affiliation{Laboratory for Terahertz $\&$ Terascale Electronics (LATTE), Department of Electrical and Computer Engineering, University 
of California-Riverside, Riverside, CA, 92521, USA}
\author{Sohee Kwon}
	\affiliation{Laboratory for Terahertz $\&$ Terascale Electronics (LATTE), Department of Electrical and Computer Engineering, University 
of California-Riverside, Riverside, CA, 92521, USA}
\author{George J. de Coster}
	\affiliation{DEVCOM Army Research Laboratory, 2800 Powder Mill Rd, Adelphi, MD, 20783, USA}
\author{Roger K. Lake}
	\email{rlake@ece.ucr.edu}
	\affiliation{Laboratory for Terahertz $\&$ Terascale Electronics (LATTE), Department of Electrical and Computer Engineering, University 
of California-Riverside, Riverside, CA, 92521, USA}
\author{Mahesh R. Neupane}
	\email{mahesh.r.neupane.civ@army.mil}
	\affiliation{Laboratory for Terahertz $\&$ Terascale Electronics (LATTE), Department of Electrical and Computer Engineering, University 
of California-Riverside, Riverside, CA, 92521, USA}
	\affiliation{DEVCOM Army Research Laboratory, 2800 Powder Mill Rd, Adelphi, MD, 20783, USA}

\begin{abstract}
Two dimensional chromium ditelluride (CrTe$_2$) is a promising ferromagnetic layered material that exhibits long-range ferromagnetic 
ordering in the monolayer limit. 
The formation energies of the different possible structural phases (1T, 1H, 2H)
calculated from density functional theory (DFT)
show that the 1T phase is the ground state, and the
energetic transition barriers between the phases, calculated by the nudged elastic band method,
are large, on the order of 0.5 eV. 
The self-consistent Hubbard $U$ correction parameters are calculated for all the phases of CrTe$_2$. 
The calculated magnetic moment of 1T-CrTe$_2$ with $\geq 2$ layers lies in the plane, 
whereas the magnetic moment of a monolayer is out-of-plane. 
Band filling and tensile bi-axial strain cause the magnetic moment of a monolayer to switch from out-of-plane to in-plane,
and compressive bi-axial strain in a bilayer causes the magnetic moment to
switch from in-plane to out-of-plane. 
The magnetic anisotropy is shown to originate from the large spin orbit coupling (SOC) of the Te atoms
and the anisotropy of the exchange coupling constants $J_{xy}$ and $J_z$ in an XXZ type Hamiltonian.
Renormalized spin wave theory using experimental values for the magnetic anisotropy energy
and Curie temperatures provides a range of values for the nearest neighbor exchange coupling.
\end{abstract}

\maketitle
\end{CJK*}
\section{Introduction}
The recent discovery of monolayer two-dimensional (2D) ferromagnetic (FM) material 
\cite{2017_CrI3_XXu_Nat,2017_Cr2Ge2Te6_FM_Louie_Xia_Zhang_Nat},
the compatibility of 2D FM materials with other 2D materials, 
and their susceptibility to external control of their magnetic properties
have made 2D FM materials a topic of high current interest.
For example, the magnetic anisotropy can be controlled by applying an external electric 
field \cite{park2020effects}, strain \cite{PhysRevB.98.144411}, and band filling \cite{PhysRevMaterials.1.074408}. 
The ground state magnetic ordering can be switched among ferromagnetic (FM), anti-ferromagnetic (AFM), collinear, 
and noncollinear by stacking pattern \cite{D1NR02480A}, strain \cite{PhysRevB.104.064416}, and electric field 
\cite{PhysRevLett.124.067602,2020_E_Tunable_FM_APL}. 
Moreover, the formation of heterostructures with other 2D materials,
breaks time reversal symmetry, which can be exploited for valleytronics 
\cite{2018_WSe2_CrI3_Valley_Splitting_XXu_NLett}
or the creation of a Chern insulator \cite{2021_MTI_Review_AdvMat}.

A relatively new class of layered magnetic materials such as CrTe$_2$, CrI$_3$ and CrGeTe$_3$ have extended the applicability of the 
layered materials in the field of spintronics \cite{och2021synthesis}. 
One material of particular interest is CrTe$_2$ in which Cr hexagonal planes are sandwiched 
by Te layers.
Several studies \cite{ataca2012stable,guo2014tuning} 
suggested the non-magnetic 2H phase was the ground state, 
whereas recent studies all find the 
1T phase to be the ground state 
\cite{freitas2015ferromagnetism,
sun2020room,
zhang2021room,
Meng2021,
2021_CrTe2_AIPAdv,
bastos2019ab}. 
\crte has one of the highest Curie temperatures among the 2D magnetic materials.
The discovery that
bulk 1T-CrTe$_2$ is a layered metallic ferromagnet with a Curie temperature of $\sim 310$ K \cite{freitas2015ferromagnetism}, 
led to a number of further studies.
Mechanical exfoliation of \crte with either h-BN or Pt encapsulation in a glove box
produced samples in which easy-plane ferromagnetism was maintained in thin-films down to $\sim 8$ nm 
while maintaining a Curie temperature above 300K \cite{sun2020room}.
This study also showed that CrTe$_2$ rapidly oxidizes in ambient conditions
and that the pristine Raman peaks at 100 cm$^{-1}$ and 134 cm$^{-1}$ shift to 
125 cm$^{-1}$ and 145 cm$^{-1}$ after a few hours in air \cite{sun2020room}.
A number of studies of epitaxial grown material quickly followed.
Thin film \crte was grown by molecular beam epitaxy (MBE) on bilayer graphene (BLG)/SiC and capped with a 5 nm Te layer
to prevent oxidation \cite{zhang2021room}.
Ultrathin films ($\leq 7$ monolayers (ML)) posessed perpendicular magnetic anisotropy (PMA)
with $T_c$ dropping from 300 K for thicker films down to 200 K for a monolayer. 
A large PMA constant of $K_u = 5.63 \times 10^6$ erg/cm$^3$ was measured for a 7 ML film.
In a separate work, 
this value of $K_u$ was also found for 80 nm thick films of Cr$_{1.3}$Te$_2$ \cite{2020_Tailoring_Mag_Okada_PRMat}.
In thin films of \crte grown by chemical vapor deposition (CVD) on SiO$_2$, 
the magnetic easy axis changed from in-plane to perpendicular
as the thickness was reduced below approximately 10 nm ($\approx 17$ MLs) \cite{Meng2021}.
Reflectance magneto circular dichroism measurements showed that
$T_c$ increased from approximately 165 K to 212 K as the film thickness decreased
from 48 nm to 7.5 nm. 
This last trend of {\em increasing} $T_c$ with {\em decreasing} 
film thickness is unique to these samples and experiments.
The majority of the data in this study was taken from oxidized samples based on the Raman peaks
at 123 cm$^{-1}$ and 143 cm$^{-1}$,
however a comparison was made between samples with and without h-BN encapsulation; the
values for $T_c$ remained essentially the same, and both sets of films exhibited strong PMA \cite{Meng2021}.
The authors theoretically found that the sign of the magnetic anisotropy energy (MAE) in
ML \crte switches from in-plane to out-of-plane
with increasing magnitude of the on-site Coulomb potential ($U$), 
with switching occurring at $U \sim 3.2$ eV;
and they discuss the possibility
that thinner samples provide less screening, larger electrostatic
interaction with the substrate, larger values of $U$, and thus PMA \cite{Meng2021}.
MBE grown \crte on (111) GaAs exhibited a Curie temperature that dropped from $T_c = 205$ K for a 35 ML
film to 191 K for a 4 ML film, and, unique to these samples, all thicknesses exhibited PMA \cite{2021_CrTe2_AIPAdv}. 
No information on a capping layer or other protection from oxidation was provided \cite{2021_CrTe2_AIPAdv}. 
A most recent study of MBE grown \crte on BLG/SiC found that ML \crte had a zigzag AFM (z-AFM) ground state
accompanied by a $2 \times 1$ reconstruction of the lattice resulting from 
relatively large substrate induced strain (-5\% along $a_1$ and +3\% along $a_2$) \cite{2022_CrTe2_spin_map_AFM_NComm}.
%

The intense interest in \crte also motivated many theoretical investigations
based on density functional theory calculations.
Calculations using the Perdew-Burke-Emzerhof (PBE) functional \cite{perdew1996generalized} 
without a Hubbard U correction or spin orbit coupling found that
$1$\% compressive strain caused ML \crte to
transition from an FM to an AFM ground state \cite{2015_CrTe2_Strain_Cntl}. 
Simulations of ML \crte with the all electron code WIEN2k \cite{WIEN2K_2003}
using the PBE functional
found imaginary modes in the phonon spectrum which were removed
in a $\sqrt{3} \times \sqrt{3}$ charge density wave (CDW) state \cite{otero2020controlled}.
In both the normal and CDW phase, tensile strain was required to obtain PMA,
and
the magnetic anisotropy swtitched from in-plane to out-of-plane
for a lattice constant of $\gtrsim 3.8$ {\AA} in the CDW phase and $\gtrsim 3.86$ {\AA}
in the normal phase \cite{otero2020controlled}.
PBE+U calculations, with $U = 2$ eV, found a stable phonon spectrum for ML \crte 
and in-plane FM magnetization \cite{2021_Tunable_AH_CrTe2_PRB}.
The finding of in-plane magnetization results from the use of the value $U=2$ eV \cite{Meng2021}.
PBE level calculations without a Hubbard $U$ correction found an AFM ground state for ML \crte, 
and a reduction of the lattice constant from 3.79 {\AA} in the bulk to 3.68 {\AA} in ML
\cite{2021_Half_Metal_CrTe2_PhysLettA}.
The ML AFM ground state was attributed to the reduction of the lattice constant.
The thickness dependence of the magnetization of \crte was investigated \cite{2021_thickness_dep_mag_crte2_JPCLett} 
using the opt-B86b-vdW functional \cite{opt-B86b-vdW_Klimes_2011} 
implemented in VASP \cite{kresse1996efficient, kresse1993ab}.
The ML ground state was found to be z-AFM
with a corresponding reduction of the in-plane lattice constant 
from $\sim 3.8$ {\AA} for bulk to $\sim 3.57$ {\AA} for ML \cite{2021_thickness_dep_mag_crte2_JPCLett}.
The FM ML CDW ground state \cite{otero2020controlled} was found to be higher in energy than the z-AFM state.
The results are qualitatively similar to those of Ref. \cite{2021_Half_Metal_CrTe2_PhysLettA}.
AFM interlayer coupling was found in 2 through 4 MLs, and FM interlayer coupling for 5 MLs or more 
\cite{2021_thickness_dep_mag_crte2_JPCLett}.
PBE-D3+($U\!=\!2$ eV) calculations of bilayer \crte
found a g-type AFM ground state with both intra-layer and inter-layer AFM coupling \cite{2021_BL_CrTe2_APL}.
Compressive strain greater than 4\% caused the interlayer coupling to become FM while the intra-layer coupling
remained AFM.

CrI$_3$ is another 2D magnetic material with many similarities to CrTe$_2$. 
The Cr ion is in octahedral coordination with the I anions resulting in the same
$e_g$, $t_{2g}$ crystal field splitting and superexchange coupling through the Cr-I-Cr bonds
at near $90^\circ$ bond angles.
The origin of the large out-of-plane magnetic anisotropy in CrI$_3$ has been investigated in detail
\cite{lado2017origin,kim2019giant}.
It was found that the MAE is primarily from the SOC on the I atoms, and, therefore,
anisotropic superexchange is the source of the magnetic anisotropy \cite{lado2017origin}.  
Further investigation found that the MAE was very sensitive to the deviation of the 
dihedral angle $\theta_D$ between the plane
formed by the Cr-I-Cr bonds and 
a vertical plane through the Cr-Cr pair \cite{kim2019giant}, which
is a measure of the trigonal distortion of the edge-sharing CrTe$_6$ octahedra. 
In the undistorted octahedron, the dihedral angle $\theta_{O_h} \approx 35.3^\circ$, 
and the deviation is defined as
$\delta \theta \equiv \theta_{O_h} - \theta_D$. 
In CrI$_3$, positive values for $\delta \theta_D$ resulted in out-of-plane magnetic 
anisotropy and negative values resulted in in-plane magnetic anisotropy.
%

The variety of different and contradicting experimental data for \crte originating from
different growth conditions and substrates 
indicates a sensitivity of the thin layer material to external perturbations such as strain, band filling
and screening.
The variety of different and contradictory theoretical predictions resulting from different models and, particularly,
from the use of different values of $U$ possibly indicate a sensitivity to screening, which is affected
by different environments as discussed in \cite{Meng2021}.
In few monolayer films, 
both the interlayer magnetic coupling and the sign of the magnetic anisotropy are affected 
in incompatible ways by the value of $U$.
For few layer films, small $U$ values give, what appears to be at this time, the experimentally correct sign 
of the interlayer magnetic coupling coupling (i.e. FM),
but the incorrect sign for the magnetic anisotropy (i.e. prediction of easy-plane magnetic anisotropy).
Conversely, larger values of $U$ predict the correct magnetic anisotropy (PMA), but the incorrect interlayer 
magnetic coupling (i.e. AFM).
Thus, to address the question of the magnetic anisotropy in a monolayer, 
a value for $U$ must be chosen that reproduces the observed
magnetic anisotropy, which, experimentally, is found to be out-of-plane.
%

In this work,  
we first quantify the energy differences and energy barriers separating the different
crystallographic phases: 1T, 1H, and 2H.
%
%
We then focus on the magnetic anisotropy of bilayer and monolayer \crte 
and understand how it is affected by strain and band filling.
We investigate the source of the magnetic anisotropy originating from the large SOC of the Te atoms.
Based on the insights gained from prior work on CrI$_3$ \cite{lado2017origin,kim2019giant}, 
we analyze the SOC matrix elements and distortion of dihedral angle, and their relationships
to the sign of the MAE.
Finite temperature long range magnetic order in 2D monolayer \crte is subject to
the Mermin-Wagner theorem \cite{1966_Mermin_Wagner_PRL}.
As such, an energy gap is required
in the magnon excitation spectrum to prevent the magnetic order from being destroyed by
thermal fluctuations.
This energy gap results from the magnetic anisotropy.
The interdependence of the
MAE, exchange coupling, and Curie temperature in ML \crte,
is analyzed using 
renormalized spin wave theory (RSWT) \cite{1962_Bloch_RWST}.
RSWT provides a mean field self-consistent calculation of the magnon mode occupation and
the average magnetic moment as a function of temperature.
Examples of RSWT applied to other 2D magnetic materials can be found in Refs\cite{2017_Cr2Ge2Te6_FM_Louie_Xia_Zhang_Nat,lado2017origin,2020_RSWT_APL}.
Finally, an inverse calculation is performed in which
the experimentally measured value for $T_c$ is used to
determine all pairs of values for the MAE and exchange coupling constants that result in $T_c$.

\begin{figure}[htpb]
	\centering
	\includegraphics[width = 0.5\textwidth]{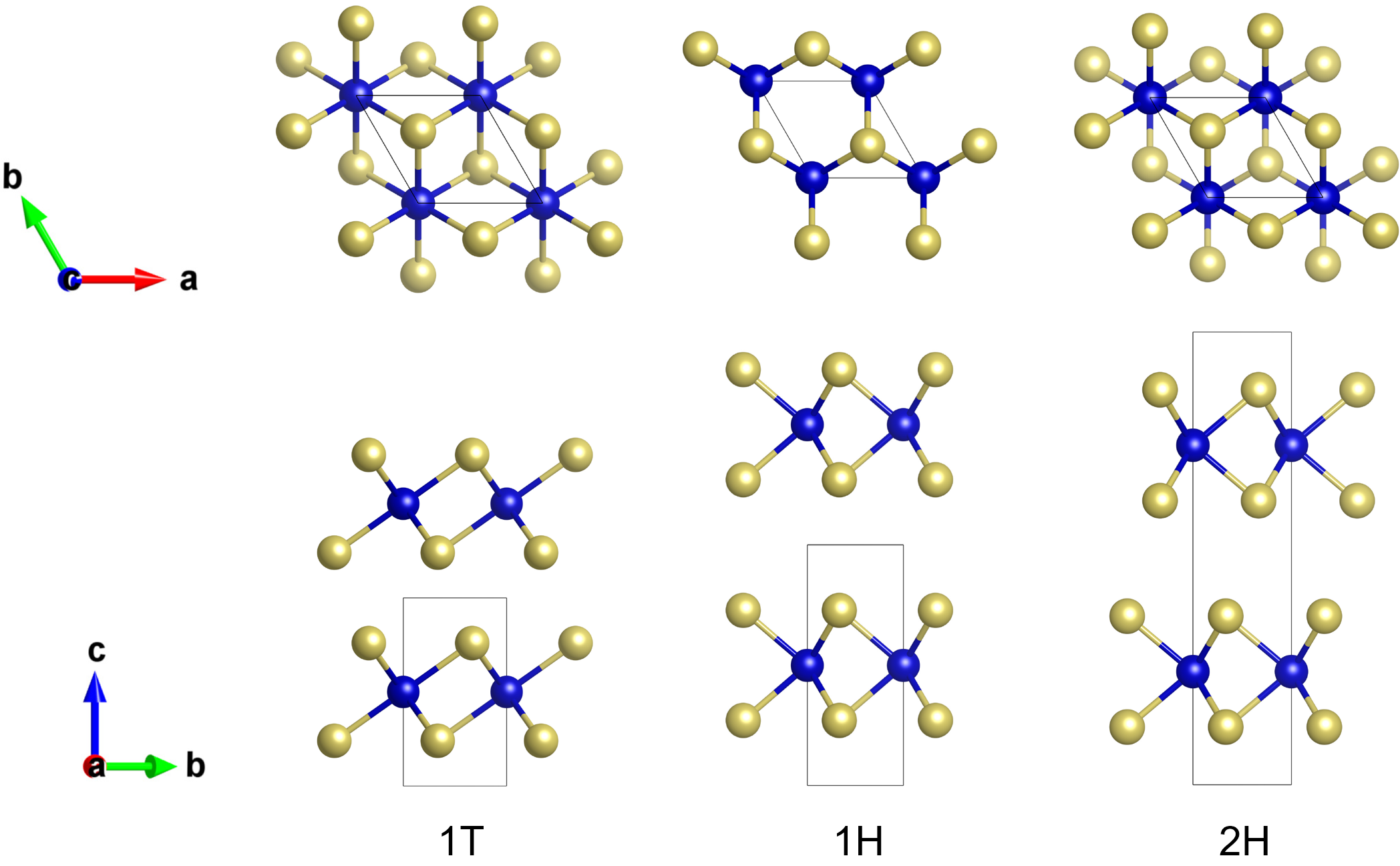}
	\caption{Top and lateral views of 1T, 1H, and 2H phases of CrTe$_2$. The unit cells are shown by the thin lines. 
	Blue and golden balls represent Cr and Te atoms, respectively.
	The 1T phase contains one formula unit (f.u.) per unit cell in a hexagonal lattice 
	belonging to the $P\bar{3}m1$ space group with 
	each Cr atom surrounded by Te atoms in octahedral coordination. 
	The 1H and 2H phases are hexagonal, trigonal prismatic, and 
	the difference between the two phases is in their interlayer stacking.
	In the 1H structure, layers are stacked directly on top of each other so that
	the 1H structure contains 1 f.u. / unit cell and belongs to the $P\bar{6}m2$ space group.
	The 2H structure contains 2 f.u. / unit cell and belongs to the $P6_{3}/mmc$ space group.}
	\label{CrTe2_phase}
\end{figure}
%

\section{Method}

The first-principle calculations use spin-polarized density functional theory (DFT) with the projector 
augmented wave (PAW) \cite{blochl1994projector, kresse1999ultrasoft} method and a plane-wave basis, 
as implemented in the Vienna ab initio simulation package (VASP) \cite{kresse1996efficient, kresse1993ab}. 
The Perdew-Burke-Enzerhof's (PBE) \cite{perdew1996generalized} version of the generalized gradient approximation (GGA) 
is used for the exchange-correlation density functional.
The vdW corrections are included with the PBE+D3 model \cite{grimme2010consistent}.
All structural relaxation calculations use the PBE+D3 level of theory.
The lattice is fully relaxed until the force on each atom is smaller than 0.001 eV/\AA. 
For finite thickness slabs, 15 {\AA} vacuum layers are added.
Energy barriers between the ground state and the metastable states of 
CrTe$_2$ are determined using the nudged elastic band (NEB) method 
\cite{henkelman2000climbing,doi:10.1021/acs.jpcc.9b07804}.

For calculation of the electronic and magnetic properties, 
the Hubbard U correction (PBE+U) \cite{PhysRevB.57.1505}. 
and spin orbit coupling (SOC) are included.
The values of the U parameter for the different phases of CrTe$_2$ are
calculated using the linear response method \cite{PhysRevB.71.035105},
and the values are given in Table \ref{U_parameter}. The details of U parameter calculation are provided in the Supplemental Material (SM) \cite{Supp}.
\begin{table}[htbp]
	\renewcommand\arraystretch{1.25}
	\caption{\label{U_parameter}U parameters of the Cr atom in CrTe$_2$ calculated from linear response method. }
	\begin{ruledtabular}
		\begin{tabular}{ccccccc}
			Phases& 1T bulk & 1H bulk & 2H bulk & 1T 1L & 1H 1L & 2H 2L\\
			\hline
			U (eV) & 5.80 & 5.59 & 5.85 & 5.92 & 5.91 &5.59 \\
		\end{tabular}
	\end{ruledtabular}
\end{table}
For all calculations of the magnetic properties of 1T-CrTe$_2$, the value of $U=5.8$ eV is used.
With $U=5.8$ eV, the magnegetic moment per formula unit of 1T-CrTe$_2$
is 3.05 $\mu_B$ for monolayer and 3.08 $\mu_B$ for bulk.
%
%
A table of calculated magnetic moments as a function of U is provided in the Supplemental Material (SM) \cite{Supp}. 
This method has been used for other 2D Cr based materials such as CrX$_3$(X = Cl, Br, I) monolayers \cite{liu2016exfoliating}.
12 valence electrons are included for Cr ($3p^{6}3d^54s^1$), and 6 
valence electrons for Te ($5s^25p^4$).
The cutoff energy is 500 eV. 
A 24$\times$24$\times$12 Monkhorst-Pack k-grid mesh \cite{monkhorst1976special} 
for bulk structures and a 
28$\times$28$\times$1 mesh for layered structures are used to ensure that the magnetic anisotropy energies are well converged.
The Gaussian smearing method is employed with a width of 0.05 eV for the structure, magnetic, 
and energy barrier calculations for insulating systems. 
For metallic systems, the Methfessel-Paxton smearing method is employed with a width of 0.05 eV.

\section{Results and Discussion}
\subsection{Ground state and energy barrier in phase transition}
%

CrTe$_2$ can potentially crystalize into various layered 
phases such as 1T, 1T$_d$, 1H, and 2H phases \cite{MX2Phases, sokolikova2020direct}, 
as illustrated in Fig. \ref{CrTe2_phase}. 
The geometry-optimized in-plane lattice constant $a$ and 
the interlayer distance $d$ for each phase is shown in Table \ref{formation}.
Among all of the possible phases, the 1T$_{d}$ phase of CrTe$_2$ in both the bulk and monolayer forms is
unstable during the structure optimization step, and hence is excluded from this study. 
Experimental values are only known for the 1T bulk phase, and our calculated values 
match well with the experimental ones of
$a$ = 3.7887 \AA \space and $c$ = 6.0955 \AA  \space \cite{freitas2015ferromagnetism}. 
%

%

To determine the energetic stability of each phase,
the formation energy $E_{form}$ is calculated from the energy difference between the 
material and isolated atoms per chemical formula, 
which is defined as
\begin{equation}
	E_{form} = E_{total} - \sum_i^n E_i 
	\label{Eform}
\end{equation}
where $E_{total}$ is the total energy of the material, $E_{i}$ is the energy of a single constituent atom, 
and $n$ is the total number of atoms in the unit cell of the material. 
A more negative $E_{form}$ corresponds to a more stable system.  
As shown in the Table. \ref{formation}, the 1T phase is the ground state for both the bulk and the monolayer 
forms.
Quantitatively, the formation energy of the 1T bulk phase is lower than 
those of the 2H and 1H phases by 0.30 and 0.40 eV, respectively.
\begin{table}[htbp]
\centering
\renewcommand\arraystretch{1.25}
\caption{\label{formation}Formation Energies $E_{form}$ (eV) and relaxed lattice constants for different phases
of CrTe$_2$ in bulk, monolayer (1L) and bilayer (2L) geometries. For the bilayer structure, $c$ 
corresponds to the interlayer Cr-Cr distance.} 
	\begin{ruledtabular}
		\begin{tabular}{cccc}
			Phases& $E_{form}$ & $a$ & $c$ \\
			\hline
			1T bulk & -10.44 &3.787 & 5.967 \\
			1T 2L &  -10.18  &3.759  & - \\
			1T 1L & -10.09  &3.692 & - \\
			1H bulk & -10.04 &3.491 & 7.493  \\
			1H 1L & -9.75  &3.646 & -\\
			2H bulk & -10.14 &3.498 & 6.951\\
			2H 2L & -9.98   &3.493 & 7.001\\
		\end{tabular}
	\end{ruledtabular}
\end{table}

\begin{figure}[htbp]
	\includegraphics[width=0.5\textwidth]{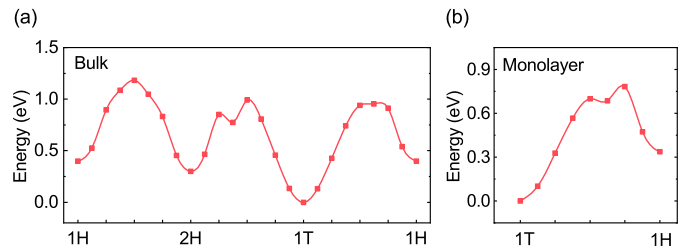}
	\caption{(a)The energy barrier between 1H, 2H, and 1T phases of bulk CrTe$_2$. (b)The energy barrier between 1T and 1H phases of 
		monolayer CrTe$_2$.}
	\label{NEB}
\end{figure}

The energetic barriers separating the ground state from the metastable states,
calculated from the NEB method, are shown in Fig. \ref{NEB} for (a) bulk and (b) monolayer.
The energies of the 1T bulk and monolayer serve as the reference energies and ere set to be 0 eV.
The energetic barriers for the bulk phase transitions from 1T to 2H and 1H are 0.99 eV and 0.95 eV, respectively.
The energetic barrier for the monolayer transition from 1T to 1H is 0.78 eV.
The large magnitudes of energy barriers separating the 1T phase from the other metastable phases
combined with the large energy differences of the ground states, indicate that the 1T phase, in both bulk
and monolayer forms, should be very stable, and transitions to other phases difficult to achieve.

To verify the stability of 1T phase monolayer, the phonon spectrum is calculated using 
different $U$ parameters as shown in the Supplemental Material (SM) \cite{Supp}. 
As found previously \cite{2021_Tunable_AH_CrTe2_PRB}, the imaginary modes vanish with the inclusion of a non-zero 
Hubbard $U$ parameter. 

\subsection{The magnetic anisotropy of layered and bulk 1T-CrTe$_2$}
The magnetic anisotropy energy plays a crucial role in the stability of the magnetic ordering
in low dimensional materials,
and there is great interest in controlling it with externally applied fields and strain. 
We therefore investigate the sensitivity of the MAE to strain and band filling in both few-layer and bulk 1T-CrTe$_2$. 
Since, the energy differences and energy barriers between the 1T phase and the other phases are large,
we only consider the magnetic properties of the 1T phase. 

The MAE ($\Delta_{\rm MA}$) is defined as the energy difference between the total energies $E_{\rm total}$ 
when the magnetization $\mv$ lies along the $x$ axis or the $z$ axis, i.e. 
\begin{equation}
\Delta_{\rm MA} = E_{\rm total}(\mv \| \xh) - E_{\rm total}(\mv \| \zh) . 
\label{eq:MAE_def}
\end{equation}
As shown in the Table \ref{MAE}, in the FM ground state, 
the magnetization easy axis of monolayer 1T-CrTe$_2$ is out-of-plane while the multilayer 
and bulk 1T-CrTe$_2$ have in-plane magnetic easy axes.
\begin{table}[htbp]
	\centering
	\renewcommand\arraystretch{1.25}
	\caption{\label{MAE} Magnetic anisotropy energies of 1T-CrTe$_2$ in layered and bulk forms.}
	\begin{ruledtabular}
		\begin{tabular}{ccc}
			Structure& MAE per f.u. (meV) & Easy axis\\
			\hline
			1L & 5.56 & out-of-plane\\
			2L & -4.15 & in-plane  \\
			3L & -3.61 & in-plane \\
			4L & -2.88 & in-plane \\
			5L & -3.37 & in-plane \\
			6L & -3.29 & in-plane \\
			Bulk & -3.22 & in-plane\\
		\end{tabular}
	\end{ruledtabular}
\end{table}
\begin{figure}[htb]
	\centering
	\includegraphics[width=8cm]{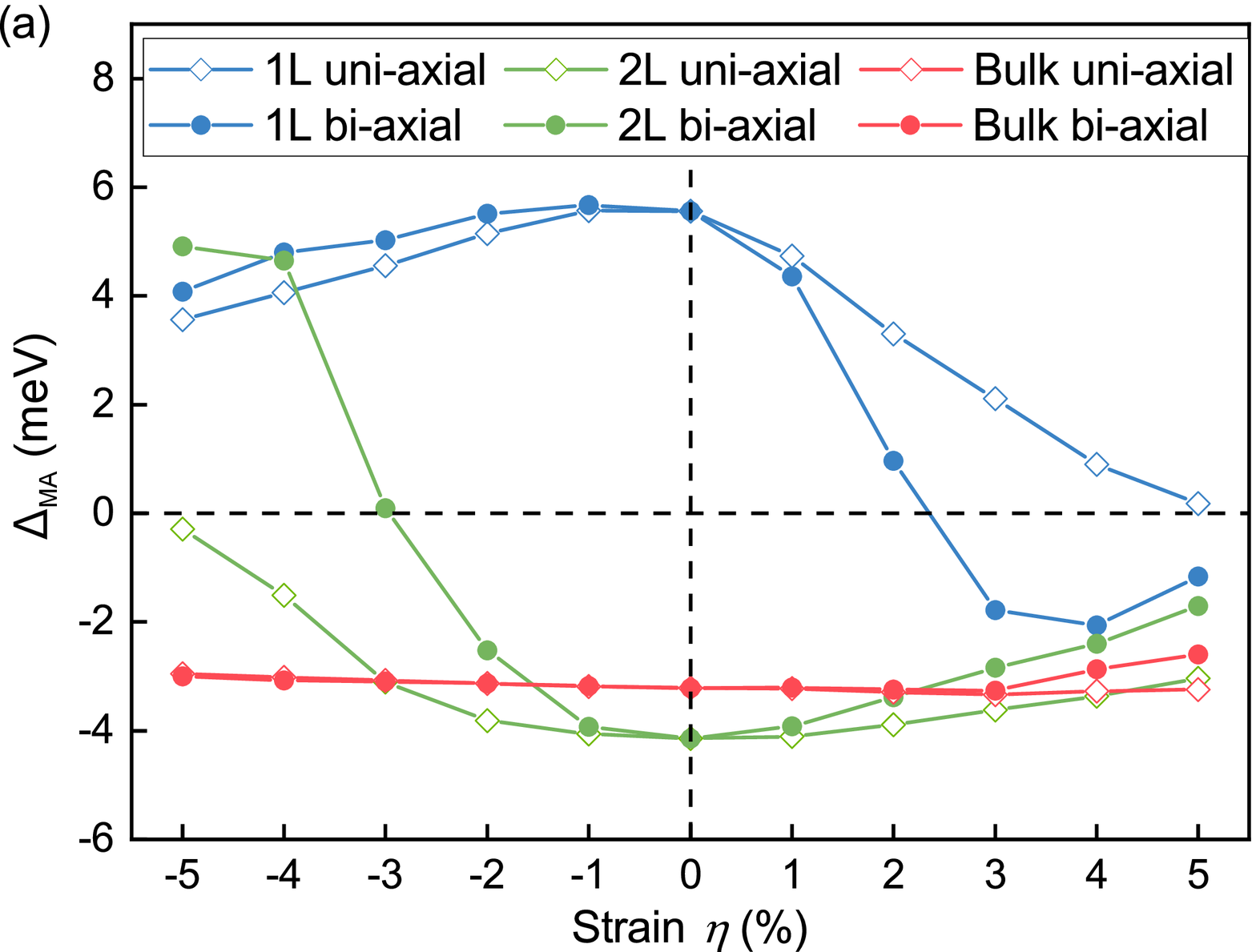}

	\includegraphics[width=8cm]{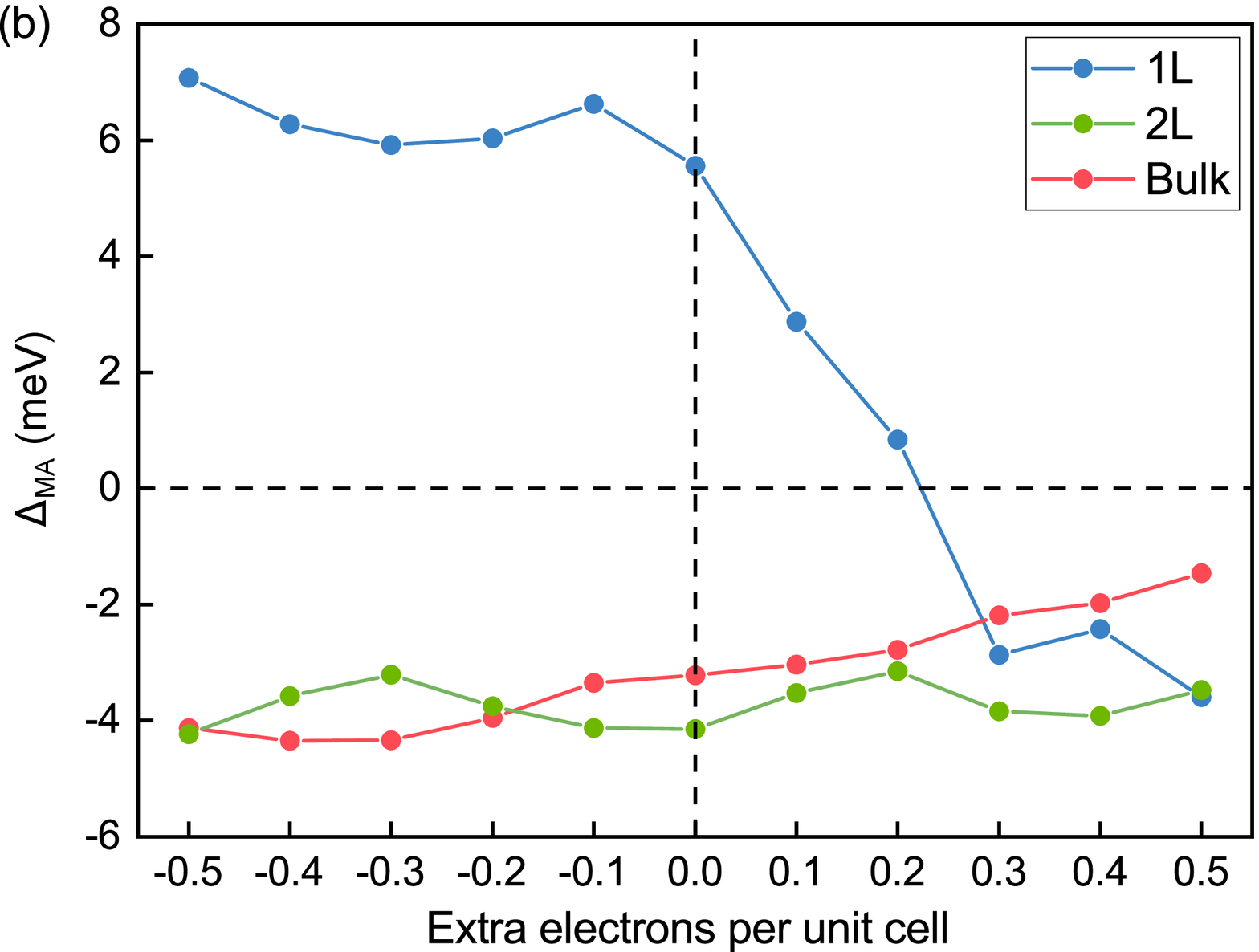}
	\caption{The MAE as a function of (a) strain and (b) band filling of 1L, 2L, and bulk 1T-CrTe$_2$}
	\label{Strain_MAE}
\end{figure}

Device applications require external control of the MAE, so we therefore consider
the effects of strain and band filling in monolayer, bilayer, and bulk 1T-CrTe$_2$.
Bi-axial strain is simultaneously stretching or compressing the in-plane $x$-axis and $y$-axis of the crystal.
As shown in the Fig. \ref{Strain_MAE}(a) 
the MAE of a monolayer is sensitive to tensile bi-axial strain, 
and the MAE of a bilayer is sensitive to compressive bi-axial strain. 
The easy axis of monolayer 1T-CrTe$_2$ switches from out-of-plane ($z$-axis) 
to in-plane ($x$-axis) at 2.3$\%$ bi-axial tensile strain.
The easy axis of bilayer 1T-CrTe$_2$
switches from in-plane ($x$-axis) to out-of-plane ($z$-axis) at 3$\%$ bi-axial compressive strain. 
The MAE of the bulk structure is relatively insensitive to the applied uni-axial or bi-axial strain. 
As shown in Fig. \ref{Strain_MAE}(b), 
band filling also switches the magnetic moment of monolayer of 1T-CrTe$_2$ 
from out-of-plane ($z$-axis) to in-plane ($x$-axis). 
The sign of the MAE switches
at a filling of 0.22 electrons per unit cell, 
corresponding to a sheet carrier concentration of $n_s = 1.9 \times 10^{14}$ cm$^{-2}$. 
\begin{figure}[tb]
	\centering
	\includegraphics[width=8cm]{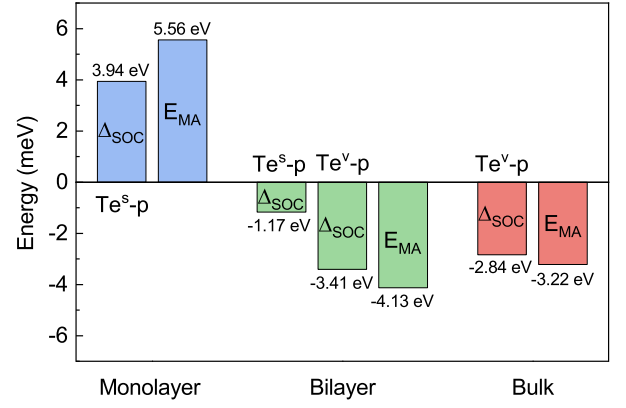}
	\caption{
		Magnetic anisotropy energy $\Delta_{\rm MA}$ (per f.u.) and difference in SOC energies $\Delta_{\rm SOC}$ (per f.u.) 
		of Te-5p orbitals between the $x$ (in-plane) and $z$ (out-of-plane) magnetization orientations. 
		Te$^s$ and Te$^v$ denote the Te atoms on the surface and at the vdW gap of the bilayer, respectively. 
	}
	\label{Source_MAE}
\end{figure}

To obtain insight into the source of the magnetic anisotropy in 1T-CrTe$_2$, 
we consider the SOC matrix elements of Cr-3d and Te-5p orbitals. 
The Cr $d$-orbitals' contributions to SOC matrix elements are negligible in comparison with 
those of the Te $p$-orbitals, so they will be ignored.
We abbreviate the $p$-orbital matrix elements of the SOC term in the Hamiltonian as $\braket{p_i}{p_j}$.
Similar to the definition of the MAE in Eq. (\ref{eq:MAE_def}), we define 
\begin{equation}
\Delta_{\rm SOC} = E_{\rm SOC}(\mv \| \xh) - E_{\rm SOC}(\mv \| \zh)
\label{eq:Delta_SOC}
\end{equation}
where $E_{\rm SOC}$ is the energy associated with the SOC matrix elements.
In Fig. \ref{Source_MAE}, $E_{\rm SOC}(\mv \| \xh,\zh)$ is calculated from
the sum of the SOC matrix elements, i.e. 
$E_{\rm SOC}(\mv \| \xh,\zh) = \left. ( \braket{p_y}{p_x} + \braket{p_y}{p_z} + \braket{p_x}{p_z} ) \right|_{\mv \| \xh,\zh}$,
and the difference $\Delta_{\rm SOC}$ is plotted.
Fig. \ref{Source_MAE} shows $\Delta_{\rm MA}$ and $\Delta_{\rm SOC}$ for monolayer, bilayer and bulk 1T-CrTe$_2$.
It is clear that the difference in the SOC energy $\Delta_{\rm SOC}$ 
tracks both the magnitude and sign of the MAE, $\Delta_{\rm MA}$. 
In the bilayer structure, the Te atoms on the outer surfaces (Te$^s$) and the ones adjacent to the vdW gap (Te$^v$)
are in different chemical environments, and thus they contribute different amounts to the total MAE. 
\begin{figure}[tb]
\centering
\subfloat{\includegraphics[width=8cm]{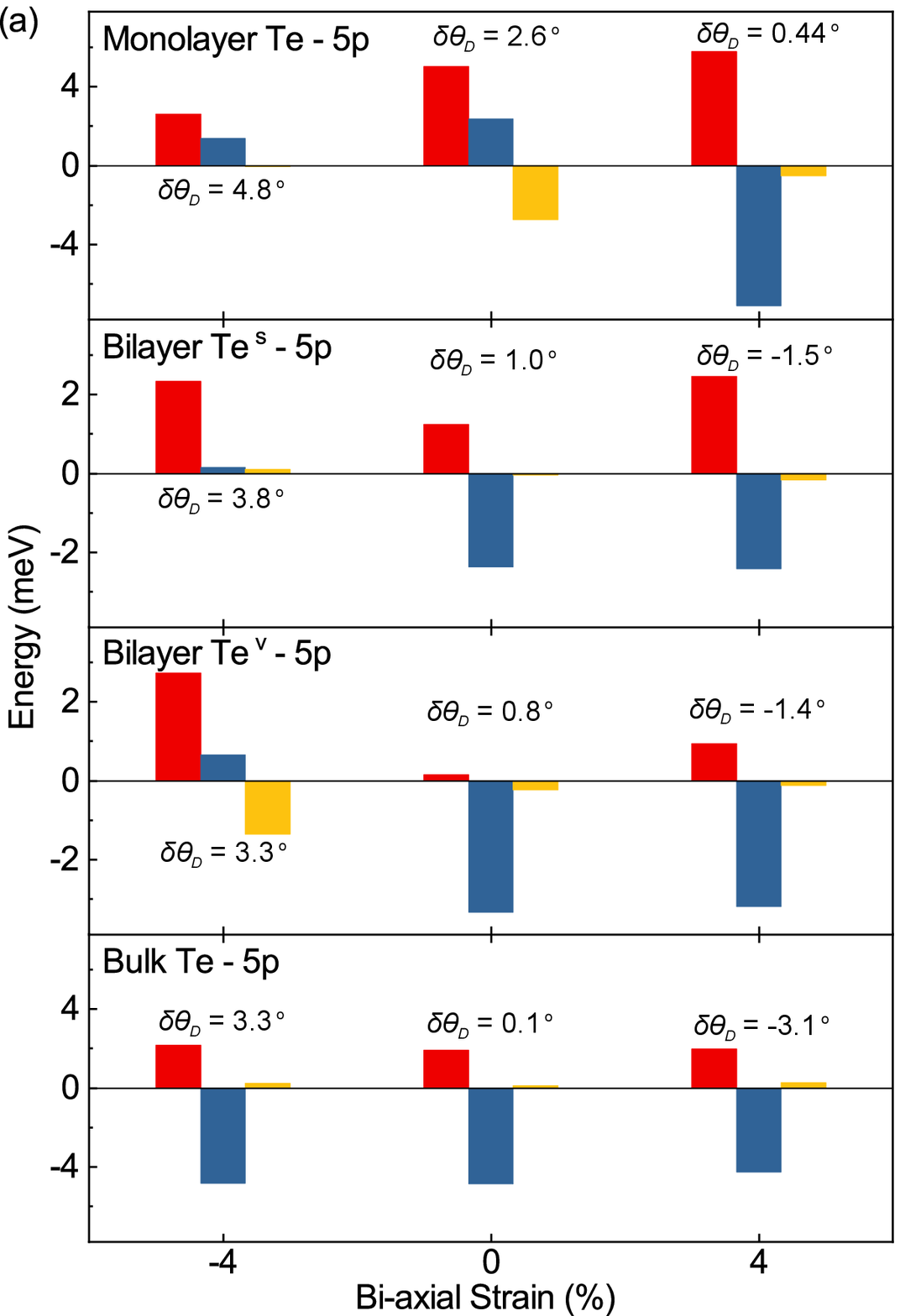}}\\
\subfloat{\includegraphics[width=8cm]{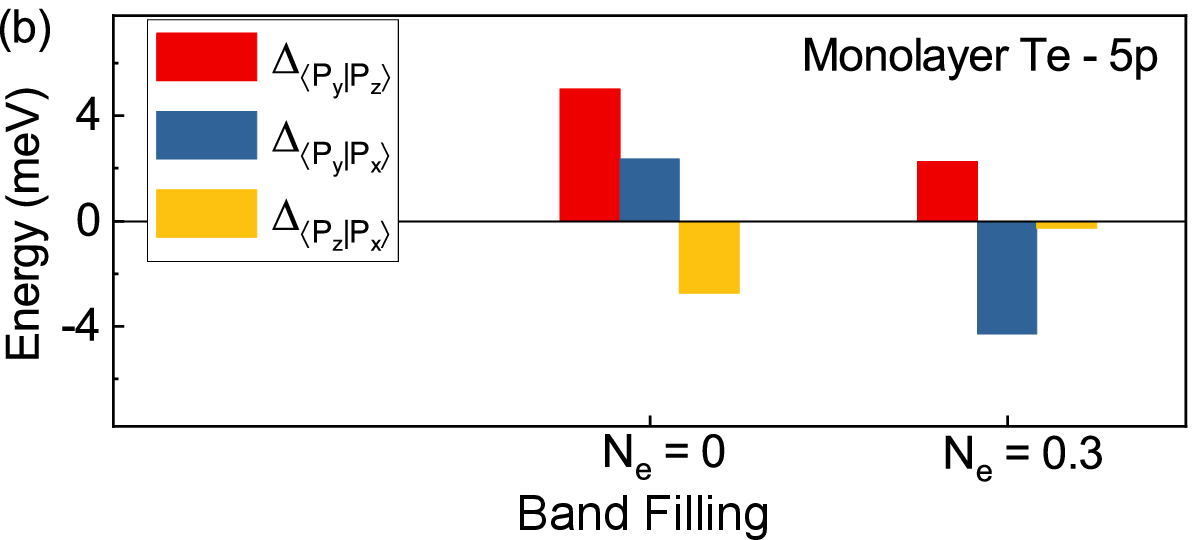}}
\caption{
Difference in SOC matrix elements $\Delta_{\braket{p_i}{p_j}}$ (per atom) of the Te-5p orbitals.
(a) $\Delta_{\braket{p_i}{p_j}}$ of 1L, 2L, and bulk 1T-CrTe$_2$ versus strain. 
At each strain, the values for $\delta \theta_D$ are also shown.
Positive and negative values of strain correspond to tensile and compressive strain, respectively.
For the bilayer, values for Te atoms at the van der Waals gap (Te$^v$) and Te atoms at the free surface
Te$^s$ are shown.
(b) $\Delta_{\braket{p_i}{p_j}}$ of 1L 1T-CrTe$_2$ versus filling. 
The legend is shown at left.
}
\label{SOC_strain_band_filling}
\end{figure}
\begin{figure}
	\centering
	\includegraphics[width=2in]{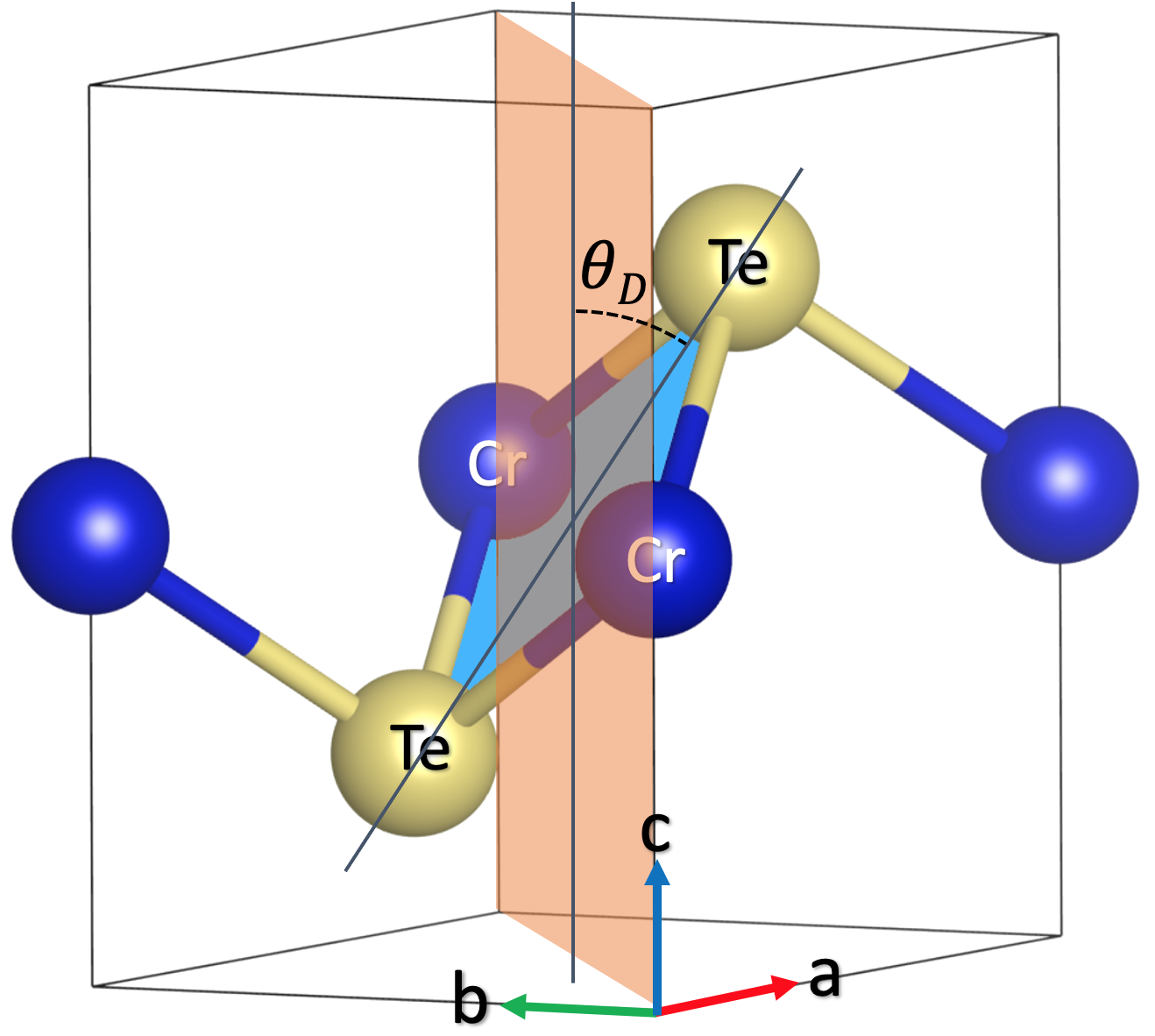}
	\caption{Illustration of the dihedral angle $\theta_D$ between the plane (blue) formed by a Cr-Te-Cr 
bonds and the perpendicular plane through the Cr-Cr pair (orange).}
	\label{fig:dihedral_angle}
\end{figure}

The changes in individual SOC matrix elements with different magnetization directions are shown in 
Fig. \ref{SOC_strain_band_filling}.
Here, 
$\Delta_{\braket{p_i}{p_j}} = \left. \braket{p_i}{p_j} \right|_{\mv \| \xh} - \left. \braket{p_i}{p_j} \right|_{\mv \| \zh}$.
In the FM ground state (zero strain) of monolayer \crte, 
$\langle p_y|p_z \rangle$ contributes the most to $\Delta_{\rm SOC}$, 
while in the FM ground states of bilayer and bulk, the dominant matrix element is $\langle p_y|p_x \rangle$.
A dominant $\Delta_{\langle p_y|p_z \rangle}$ matrix element anisotropy coincides with an out-of-plane
easy axis,
and a dominant $\Delta_{\langle p_y|p_x \rangle}$ matrix element anisotropy coincides with an
in-plane magnetic easy axis.
%
%

Fig. \ref{SOC_strain_band_filling}(a) also shows the effect of strain on the dihedral angle 
$\theta_D$
between the Cr-Te-Cr plane and a vertical plane through the Cr-Cr pair illustrated in 
Fig. \ref{fig:dihedral_angle}.
Positive values of $\delta \theta_D = \theta_{O_h} - \theta_D$ correspond to the Cr-Te-Cr plane
becoming more vertical.
For the monolayer and bilayer, 
an out-of-plane easy axis 
occurs at more positive values of 
$\delta \theta_D$,
which is qualitatively consistent with the results for CrI$_3$ described in Ref. \cite{kim2019giant},
although the dependence is far from linear.
For the monolayer
In equilibrium, $\delta \theta_D = 2.6^\circ$ is relatively large and positive, 
the $\Delta_{\langle p_y|p_z \rangle}$ matrix element anisotropy
is dominant, and the easy axis is out of plane.
For the bilayer in equilibrium, 
two values of $\delta \theta_D$ are given, one for the Te atom at the surface ($1.0^\circ$) and one for the
Te atom at the van der Waals gap ($0.8^\circ$). 
The angles are similar, $\delta \theta_D \sim 1^\circ$, the matrix element anisotropy is 
dominated by $\Delta_{\braket{p_y}{p_x}}$, and the easy axis is in plane.
As compressive bi-axial strain is applied to the bilayer, $\delta \theta_D$ becomes more positive,
the SOC matrix element anisotropy $\Delta_{\langle p_y|p_z \rangle}$ becomes dominant,
and the easy plane rotates from in-plane to out-of-plane.
For the bulk, compressive strain increases $\delta \theta_D$ to $3.3^\circ$, however in the bulk,
the anisotropy of the SOC matrix elements and the magnetic anisotropy are insensitive to strain
and the dihedral angle.

In terms of percent change, 
strain has the largest effect on the dihedral angle $\theta_D$, the second largest effect on
the Cr-Te-Cr bond angles, and minimal effect on the bond lengths. 
The distortion produced by in-plane strain 
or a reduction of the equilibrium in-plane lattice constant
is primarily absorbed by the dihedral angles and bond angles.
The decrease in the dihedral angle $\theta_D$ 
is accompanied by a reduction of the Cr-Te-Cr bond angle.
For example, the equilibrium in-plane lattice constant of a monolayer (3.692 \AA) is 2.5\% 
smaller than that of the bulk (3.787 \AA),
the Cr-Te-Cr bond angle of the ML ($86.3^\circ$) is 4\% smaller than that of the bulk ($89.9^\circ$), 
and the dihedral angle ($32.7^\circ$) is 7\% smaller than that of the bulk ($35.2^\circ$).
Even though the lattice constants of the ML are 2.5\% smaller than those of the bulk,
the Cr-Te bond lengths of the ML (2.70 \AA) are 0.7\% longer than those of the bulk (2.68 \AA),
since the Te atoms in the ML are free to move into the vacuum.
The bond angles and bond lengths of the equilibrium bilayer lie in between those of the monolayer and bulk.
The bilayer Cr-Te$^{v(s)}$ bond lengths are 2.68(2.69) \AA, the Cr-Te$^{v(s)}$-Cr bond angles are
$88.9^\circ$($88.5^\circ$), and the Cr-Te$^{v(s)}$-Cr dihedral angles are $34.5^\circ$($34.3^\circ$).
When the bilayer is compressed 4\% in-plane, 
the Cr-Te$^{v(s)}$ bond lengths are reduced by 0.6\%(0.09\%), the Cr-Te$^{v(s)}$-Cr bond angles are reduced
by 4.3\%(4.8\%), and the Cr-Te$^{v(s)}$-Cr dihedral angles are reduced by 9\%(10\%).

As shown in Fig. \ref{SOC_strain_band_filling}(b), 
band filling in the monolayer 
switches the $\langle p_y|p_x \rangle$ anisotropy
from positive to negative, and it decreases the magnitudes of the other two terms. 
The net result is that
the MAE becomes dominated by the $\Delta_{\langle p_y|p_x \rangle}$ term, and the 
easy axis switches from out-of-plane to in-plane.
In the band-filling calculation, the structure is not relaxed after charging, so all of the dihedral angles and
bond angles remain the same as in the charge neutral state.
Thus, this switching is a purely electronic effect.

\subsection{XXZ Spin Hamiltonian}

Magnetic anisotropy originating from the nonmagnetic ligand $p$ electrons is induced by the superexchange 
mechanism \cite{kim2019giant, lado2017origin, chang2020anisotropic} through the Cr-Te-Cr channel. 
Magnetic anisotropy of this kind is exchange anisotropy, 
in which the exchange coupling constants 
of the Heisenberg Hamiltonian 
depend on the directions of the magnetic moments. 
A suitable model is the XXZ Hamiltonian in which the the exchange coupling constant 
for the in-plane component of the spins $J_{xy}$ differs from the out-of-plane
component $J_z$ \cite{PhysRevB.52.310}.
This model has been shown to apply to CrI$_3$ \cite{lado2017origin}.
Other sources of anisotropy include single ion anisotropy and dipolar coupling.
The effect of dipolar coupling is known to be small, so 
below, we consider a XXZ type Hamiltonian for the energy per unit cell
of monolayer 1T-CrTe$_2$
and also include a single-ion anisotropy
term and an external magnetic field directed in the $\pm z$ direction (along the $c$ axis),
\begin{align}
H &= -J_{xy} \tfrac{1}{N} \sum_{i \ne j}(S_{i}^{x}S_{j}^{x} + S_{i}^{y}S_{j}^{y})  - J_{z} \tfrac{1}{N} \sum_{i \ne j}(S_{i}^{z}S_{j}^{z}) 
\nonumber \\
& - K_u \tfrac{1}{N} \sum_j (S_j^z)^2 + g \mu_B B_z \tfrac{1}{N}\sum_j S_j^z .
\label{eq:Anisotropic_Heisenberg_H}
\end{align}
Since 1T-CrTe$_2$ is ferromagnetic, 
the exchange coupling constants $J_{xy}$ and $J_z$ are positive. 
For perpendicular magnetic anisotropy in a monolayer, $K_u$ is positive.
The spin magnetic moment $M_j^z = - g \mu_B S_j^z$, so that the last term is
$-\frac{1}{N}\sum_j {\bf M}_j \cdot {\bf B}$ with ${\bf B}$ directed along the $\pm z$ direction.
Exchange coupling is included for nearest neighbor Cr ions.

The magnon dispersion is determined by first
performing the Holstein-Primakoff transformation \cite{Holstein_Primakoff1940}
defined by the operator substitution
$S_j^z = S - \hat{n}_j$, 
where $\hat{n}_j = a_j^\dagger a_j$ is the magnon number operator.  
One can then show that the spin ladder operators, $S_j^+ = S^x_j + i S^y_j$ and 
$S^-_j = S^x_j - i S^y_j$,
are given by
$S^+_j = \sqrt{2S} \sqrt{1 - \frac{\hat{n}_j}{2S}} \; a_j$,
and
$S^-_j = \sqrt{2S} a^\dagger_j \sqrt{1-\frac{\hat{n}_j}{2S}}$.
At low temperatures such that $\langle \hat{n}_j \rangle \ll S$, 
one expands out the square root terms to first order in $\hat{n}_j$ to obtain
\begin{align}
\hat{S}_j^z & = S - \nh_j \nonumber \\
\hat{S}_j^+ & \approx \sqrt{2S} \left( 1 - \frac{\nh_j}{4S} \right) a_j \nonumber \\
\hat{S}_j^- & \approx \sqrt{2S} \ajd \left( 1 - \frac{\nh_j}{4S} \right) 
\label{eq:HP_1st_order}
\end{align}
These are the equations used to transform the Hamiltonian in Eq. (\ref{eq:Anisotropic_Heisenberg_H_Keff}).
Keeping terms to first order in $\hat{n}_j$,
and substituting the Fourier representation of the operators 
$a_j = \frac{1}{\sqrt{N}} \sum_\kv e^{-i\kv \cdot \Rv_j} a_\kv$ (see SM for details),
the Hamiltonian governing the magnon dynamics is
\begin{equation}
H_{m} = \tfrac{2S}{N} \sum_{\kv} \left[ J_z Z + K_u - \tfrac{g\mu_B B_z}{2S} - J_{xy} {\rm Re}\{f(\kv)\} \right] a^\dagger_\kv a_\kv ,
\label{eq:H_AH_magnon}
\end{equation}
where $Z = 6$ is the number of nearest neighbor Cr atoms,
and $f(\kv) \equiv \sum_\bdelta e^{-i\kv \cdot \bdelta} \in \mathbb{R}$ 
is the form factor resulting from the
sum over the 6 nearest Cr neighbors located at the vertices of the hexagon given
explicitly by
$f(\kv) = 2 \left[ \cos(k_xa) + 
2 \cos\left(\tfrac{k_xa}{2}\right) \cos\left(\tfrac{\sqrt{3}}{2}k_ya\right) \right]$. 
In the limit of small $ka$, the magnon energy given by Eq. (\ref{eq:H_AH_magnon}) reduces to
\begin{equation}
E_m(k) = 12S \left( J_z \! - \!  J_{xy} \right)  +  2SK_u  -  g\mu_B B_z + 3SJ_{xy} k^2 a^2.
\label{eq:E-k_mag_CrTe2_parabolic}
\end{equation}

\begin{figure}[tb]
\centering
\includegraphics[width=3.4in]{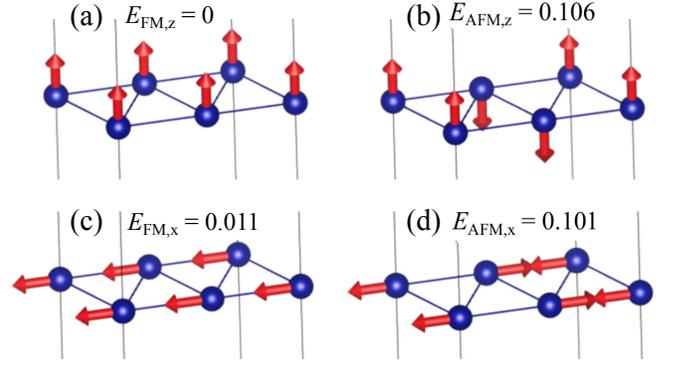}
\caption{Different spin configurations in a $2 \times 1 \times 1$ supercell,
showing only the Cr atoms,
used for determining the exchange and anisotropy parameters
in Eq. (\ref{eq:Anisotropic_Heisenberg_H}). 
The $2 \times 1$ supercell is shown with the solid black  line.
Nearest neighbors for the 2 atoms in the supercell are shown with the thinner blue lines.
The total energy (in eV) with respect to $E_{FM,z}$ 
is shown for each spin configuration.
}
\label{fig:spin_configs}
\end{figure}

The parameters $J_{xy}$, $J_z$, and $K_u$ are extracted from the DFT calculated total energies of 
structures with different spin configurations as shown in Fig. \ref{fig:spin_configs}.
From the Hamiltonian of Eq. (\ref{eq:Anisotropic_Heisenberg_H}), the total energies for each 
spin configuration for two unit cells (2 Cr atoms) are
$E_{FM,z} = 24 S^2 J_z + 2 S^2 K_u + E_0$, 
$E_{AFM,z} = -4 S^2 J_z + 2 S^2 K_u + E_0$,
$E_{FM,x} = 12 S^2 J_{xy} + E_0$, 
and 
$E_{AFM,x} = -4 S^2 J_{xy} + E_0$.
The exchange and anisotropy constants are then
\begin{align}
J_{xy}&= (E_{FM,x} - E_{AFM,x}) / (16 S^2) 
\nonumber \\
J_z&= (E_{FM,z} - E_{AFM,z})/(16 S^2)
\nonumber \\
K&= (E_{FM,z} - E_{FM,x} - 12 J_z S^2 - 12 J_{xy} S^2 )/(2 S^2) .
\label{eq:J_and_K_vals}
\end{align}
The values obtained with $U = 5.8$ eV are
$J_z = 2.93$ meV, 
$J_{xy} = 2.50$ meV, 
and $K_u = -0.0959$ meV.

Eq. (\ref{eq:E-k_mag_CrTe2_parabolic}) shows that the effective anisotropy 
governing the spin gap in the magnon dispersion 
is
\begin{equation}
K_{\rm eff} =  (J_z - J_{xy} ) Z  + K_u  = 2.49 \; {\rm (meV).}
\label{eq:Keff}
\end{equation}
We can confirm the assumption that the dipolar energy can be neglected,
since an estimate for the magnitude of the dipolar energy is 
$\frac{\mu_0}{2} M_s^2 = \frac{\mu_0}{2} (g \mu_B S)^2 V_{uc} = 41$ $\mu$eV,
where $V_{uc}$ is the volume of one unit cell. 
This is one order of magnitude smaller than the effective anisotropy energy.
With the definition of $K_{\rm eff}$, the XXZ Hamiltonian of Eq. (\ref{eq:Anisotropic_Heisenberg_H}) maps onto
a Heisenberg Hamiltonian with isotropic exchange and single-ion anisotropy as
\begin{align}
H =  \tfrac{-J}{N} \sum_{i \ne j} \Sv_i \cdot \Sv_j
-\tfrac{K_{\rm eff}}{N} \sum_j (S_j^z)^2 + g \mu_B B_z \tfrac{1}{N}\sum_j S_j^z ,
\label{eq:Anisotropic_Heisenberg_H_Keff}
\end{align}
where $J = J_{xy}$,
and the low energy magnon dispersion becomes
\begin{equation}
E_m(k) =  2SK_{\rm eff}  -  g\mu_B B_z + 3SJ k^2 a^2.
\label{eq:E-k_mag_CrTe2_parabolic_Keff}
\end{equation}

Since the effect of anisotropic exchange and single ion anisotropy enter 
the equations governing the observables of magnetic anisotropy and magnon dispersion
in exactly the same way, 
experimental measurements of effective anisotropy energies, spin-wave gaps, and
the resulting transition temperatures do not help to separate these two effects.
However, the experimental measurements of effective anisotropy and transition temperature
can shed light on the relative magnitudes of $J$ and $K_{\rm eff}$.
Two seperate experimental investigations extracted effective anisotropy constants 
from magnetization versus field curves
of $K_{\rm exp} = 5.6 \; {\rm Merg/cm}^3 = 0.26$ meV/u.c., where, for the conversion, we use the
volume of the bulk unit cell \cite{2020_Tailoring_Mag_Okada_PRMat,zhang2021room}.
The experimentally measured Curie temperatures of few layer 1T-CrTe$_2$ 
range from $T_c \sim 200 - 300$ K \cite{sun2020room,zhang2021room,Meng2021,2021_CrTe2_AIPAdv}.
From the value of $\Kexp$ and the range of values for $T_c$, 
we can extract a range of values for $J$ using renormalized spin wave 
theory (RSWT) \cite{2011_RSWT_PRB,lado2017origin,2020_RSWT_APL}.

Starting from the operator identity $S^z_j = S - \hat{n}_j$ and the saturation magnetization
per unit cell $M_s = g \mu_B S$,
the expected value of the magnetization as a function of temperature is 
\begin{equation}
M(T) = M_s - g \mu_B \tfrac{1}{N} \sum_\kv \langle \hat{n}_\kv \rangle
\label{eq:M(T)_linear}
\end{equation}
where $\langle \hat{n}_\kv \rangle = [ e^{E_m(k)/k_BT} - 1]^{-1}$ is given by
the Bose-Einstein factor.
The renormalization is included by replacing $S$ in the dispersion relation by 
$S - \tfrac{1}{N} \sum_\kv \langle \hat{n}_\kv \rangle = \frac{M(T)}{g \mu_B}$.
With this substitution, and using the expression for the low-energy dispersion (Eq. (\ref{eq:E-k_mag_CrTe2_parabolic_Keff})),
the equation for $M \equiv M(T)$ in units of $\mu_B$ becomes
\begin{equation}
M = M_s -  \tfrac{g A_{uc}}{2\pi} \!\! 
\int_0^{k_{max}}\!\!\!\!\!\!\!\!\!\! dk \: \frac{k}{e^{\tfrac{M}{g} (2 \Keff + 3 J k^2 a^2)/k_BT} - 1},
\label{eq:M(T)_linear_int}
\end{equation}
where the sum over the two dimensional wavevector is converted into an integral,
$A_{uc} = a^2 \sqrt{3}/2$ is the area of a unit cell, and $k_{max}$ is chosen
to match the area of the first Brillouin zone, i.e. $\pi k_{max}^2 = \left( \frac{4\pi}{3a} \right)^2 \sqrt{3}/2$.
Performing the integral gives
\begin{equation}
\frac{M}{g} = \frac{M_s}{g} - \frac{\sqrt{3} k_BT}{24\pi \tfrac{M}{g} J}
\ln \left[
\frac{1 - e^{-E_{\rm max}/k_BT}}{1-e^{-E_{\rm min}/k_BT}}
\right]
\label{eq:M(T)_RWST_lowE}
\end{equation}
where
$E_{\rm min} = 2 \frac{M}{g} \Keff$ and $E_{\rm max} = E_{\rm min} + \frac{8 \pi}{\sqrt{3}}\frac{M}{g}J$.

\begin{figure}[bt]
\centering
\includegraphics[width=3.4in]{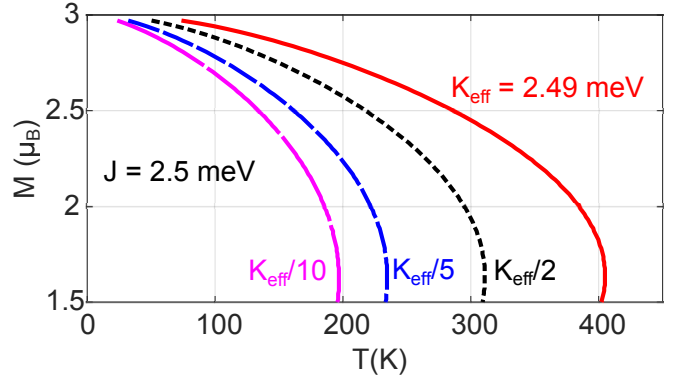}
\caption{Magnetization versus temperature with $J=2.5$ meV and
four different values of $\Keff$ as shown on the plot.
}
\label{fig:MvsT_J2.5_4Keffs}
\end{figure}

Eq. (\ref{eq:M(T)_RWST_lowE}) is solved for $M$, and $M(T)$ is plotted versus $T$ in Fig. \ref{fig:MvsT_J2.5_4Keffs}
with $J=2.5$ meV and four different values of $\Keff$.
The solid red curve with $T_c = 405$ K results from the DFT calculated parameters of 
$\Keff = 2.49$ meV and $J = 2.50$ meV.
The other curves show the effect of reducing $\Keff$.
As $\Keff$ is reduced by factors or 2, 5, and 10,
$T_c$ decreases from 405 K to 311 K, 234 K, and 197 K, respectively.
These curves illustrate the sensitivity of $T_c$ to the parameters $J$ and $\Keff$.

\begin{figure}[tb]
\centering
\includegraphics[width=3.4in]{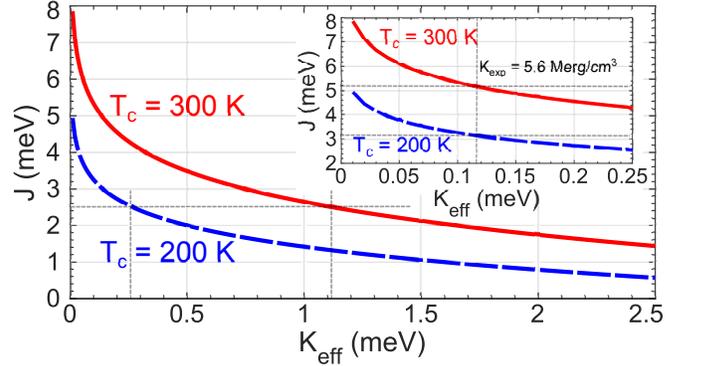}
\caption{Values of $J$ and $\Keff$ that result in $T_c = 300$ K and 200 K calculated from 
Eq. (\ref{eq:M(T)_RWST_lowE}).  Inset: Enlarged view for small $\Keff$.
To convert to values for experimentally determined anisotropy, $\Kexp = S^2 \Keff = 2.25 \Keff$ (see Supplemental Material (SM) \cite{Supp} and references \cite{zhang2021room, 1996_MJohnson_MAE_RepProgPhys} therein).
The vertical dashed line in the inset designates the experimentally measured
$\Kexp = 5.6$ Merg/cm$^3$.
}
\label{fig:Keff_J_200-300K}
\end{figure}

The pairs of parameters $J$ and $\Keff$ that result in a given value for $T_c$ form a curve in the two dimensional
$J - \Keff$ parameter space.
We solve for that curve by setting $T = T_c$ and $M = 1.65 \mu_B$ in Eq. (\ref{eq:M(T)_RWST_lowE}).
The value of $M = 1.65 \mu_B$ is chosen, since we find it to be at the point, or extremely close to the point, 
where the maximum temperature occurs in all of the 
$M(T)$ versus $T$ curves such as those shown in Fig. \ref{fig:MvsT_J2.5_4Keffs}.
The $J - \Keff$ curves showing all parameter pairs resulting in $T_c = 200$ K and $T_c = 300$ K
calculated from Eq. (\ref{eq:M(T)_RWST_lowE})
are shown in Fig. \ref{fig:Keff_J_200-300K}.

The one experimental value of $\Kexp = 5.6$ Merg/cm$^3$ is marked by the dashed vertical line on the inset.
The experimentally measured range of $T_c$ values between 200 K and 300 K
then provides a range of $J$ values between 3.1 meV and 5.2 meV.
The dashed horizontal line on the main plot shows the DFT calculated value of $J = 2.5$ meV, which
is slightly below the minimum extracted value from the experimental data.
The calculated $\Keff = (J_z-J_{xy})Z = 2.5$ meV is too large.
It is a result of the large value of magnetic anisotropy energy, $\Delta_{\rm MA} = 5.56$ meV, and the
large coordination number, $Z=6$.
We note that our values for $\Delta_{\rm MA}$ are similar to and somewhat less than those calculated in 
Ref. \cite{Meng2021} with similar values of $U$.

Care is required when comparing to other values for $\Keff$ and $J$ in the literature.
Some authors define their Heisenberg Hamiltonian as $\frac{J'}{2} \sum_{i\neq j} \Sv_i \cdot \Sv_j$ 
\cite{lado2017origin,2019_Tc_2D_Mag_Mats_PRB},
in which case, their values for $J$ must be scaled by a factor of 2 for comparison with our values.
Many authors set $S=1$ when extracting exchange constants from the DFT total energy calculations
\cite{2021_Tunable_AH_CrTe2_PRB,2021_thickness_dep_mag_crte2_JPCLett} while other authors do not
\cite{lado2017origin,2019_Tc_2D_Mag_Mats_PRB}.
Values for $J'$ calculated with $S=1$ must be scaled by $S^2$ to compare with our values for $J$, i.e.
$J = J'/S^2$.
The same also holds true for the anisotropy constant $K$.
Below, we compare to other values in the literature appropriately scaled (i.e. by a factor of $\frac{1}{2}$
or by a factor of $S^2 = \frac{9}{4}$) to match the definition of our values,
and
we also convert to our sign convention in which positive $J$ and $K$ values
correspond to FM coupling and PMA, respectively.
Based on PBE+$U=2$ eV total energy calculations of ML \crte
and fitting to a symmetric second neighbor Heisenberg model with single-ion anisotropy,
Ref. \cite{2021_Tunable_AH_CrTe2_PRB} 
found a nearest neighbor exchange interaction of $J_1 = 5.9$ meV, 
a second neighbor interaction of $J_2 = 1.1$ meV,
and an anisotropy constant of $K= -0.46$ meV.
We note the large value of $J_1$, however,
since the magnetic anisotropy is in-plane, RSWT predicts $T_c = 0$ K. 
As noted above, ML CrI$_3$ has many similarities to ML CrTe$_2$ such as 
octahedral coordination, anions with large SOC, and PMA.
PBE+$U=2.7$ eV calculations of ML CrI$_3$ found $J = 1.1$ meV and $\Keff = 0.045$ meV \cite{lado2017origin}.    
The lower values are consistent with the lower value of $T_c$ for  CrI$_3$ with respect to that of CrTe$_2$.
In general, the exchange constants straddle the ones predicted from experimental data analysed with RSWT.
The predicted anisotropy constants tend to have a stronger dependence and even change sign depending on the
value of $U$.

\section{Conclusions}
In this study, we performed a systematic first principle DFT calculations of the structural, electronic, 
and magnetic properties of various phases of CrTe$_2$. 
A comparison of the formation energies of the different phases of CrTe$_2$ show that the \crte
phase is the ground state. 
For the bulk and monolayer, 
the formation energy of the 1T phase lies 0.30 eV and 0.11 eV per formula unit below the next metastable phase, respectively.
Furthermore, NEB calculations show that the energy barriers separating the phases are large, on the order
of 0.5 eV for both the bulk and monolayer.
Based on the linear response method, the calculated $U$ value for the Cr atom in 1T-CrTe$_2$ is 5.8 eV. 
The magnetic anisotropy of 1T-CrTe$_2$ originates from the SOC of the Te atoms and the 
superexchange coupling between the Cr-3d and Te-5p orbitals.
For any number of layers ($n \ge 2$) of 1T-CrTe$_2$, the magnetic moment lies in-plane, however for a monolayer,
the magnetic moment is out-of-plane.
Band filling with a sheet carrier concentration more than $n_s = 1.5 \times 10^{14}$ cm$^{-2}$ 
or a tensile bi-axial strain of 3$\%$ can cause the magnetic easy axis of monolayer 1T-CrTe$_2$ switch from out-of-plane to in-plane. 
Compressive bi-axial strain of -3$\%$, causes
the magnetic easy axis of bilayer 1T-CrTe$_2$ to switch from in-plane to out-of-plane.
PMA is favored in structures with smaller dihedral angles consistent with the trend identified previously for CrI$_3$.
A RSWT analysis using experimental values for magnetic anisotropy and $T_c$, provides a range of expected values for
the nearest neighbor exchange constant lying between 3.1 meV and 5.2 meV for values of
$T_c$ in the range of 200 K and 300 K, respectively. 

\begin{acknowledgments}
	This work was supported in part by the U.S. Army Research Laboratory (ARL) Research Associateship Program (RAP) Cooperative 
Agreement(CA) W911NF-16-2-0008. This work used the Extreme Science and Engineering Discovery Environment 
(XSEDE) \cite{towns2014xsede}, 
which is supported by National Science Foundation Grant No. ACI-1548562 and allocation ID TG-DMR130081.
YL acknowledges helpful discussions with Prof. Ran Cheng.
\end{acknowledgments}

\providecommand{\noopsort}[1]{}\providecommand{\singleletter}[1]{#1}%
%

\pagebreak
\clearpage
\widetext
\begin{center}
\textbf{\large Supplemental Materials: Title for main text}
\end{center}
\setcounter{equation}{0}
\setcounter{figure}{0}
\setcounter{table}{0}
\setcounter{page}{1}
\makeatletter
\renewcommand{\theequation}{S\arabic{equation}}
\renewcommand{\thefigure}{S\arabic{figure}}
\renewcommand{\bibnumfmt}[1]{[S#1]}
\renewcommand{\citenumfont}[1]{S#1}

\title{Supplemental Material to ``Structural, electronic, and magnetic properties of CrTe$_2$"}
\author{Yuhang Liu}
	\email{yliu446@ucr.edu}
	\affiliation{Laboratory for Terahertz $\&$ Terascale Electronics (LATTE), Department of Electrical and Computer Engineering, University 
of California-Riverside, Riverside, CA, 92521, USA}
\author{Sohee Kwon}
	\affiliation{Laboratory for Terahertz $\&$ Terascale Electronics (LATTE), Department of Electrical and Computer Engineering, University 
of California-Riverside, Riverside, CA, 92521, USA}
\author{George J. de Coster}
	\affiliation{DEVCOM Army Research Laboratory, 2800 Powder Mill Rd, Adelphi, MD, 20783, USA}
\author{Roger K. Lake}
	\email{rlake@ece.ucr.edu}
	\affiliation{Laboratory for Terahertz $\&$ Terascale Electronics (LATTE), Department of Electrical and Computer Engineering, University 
of California-Riverside, Riverside, CA, 92521, USA}
\author{Mahesh R. Neupane}
	\email{mahesh.r.neupane.civ@army.mil}
	\affiliation{Laboratory for Terahertz $\&$ Terascale Electronics (LATTE), Department of Electrical and Computer Engineering, University 
of California-Riverside, Riverside, CA, 92521, USA}
	\affiliation{DEVCOM Army Research Laboratory, 2800 Powder Mill Rd, Adelphi, MD, 20783, USA}

\maketitle
\noindent
{\bf The calculation of Hubbard Parameters (U) for Cr in CrTe$_2$}\\

The accuracy of DFT+U calculations depends on the choice of the system dependent parameter, U. In general, the value of U parameter is 
determined empirically to match experimental structural and electronic properties of a given material. 
The values of U parameters in all different phases of CrTe$_2$ were calculated by using the linear response 
method\cite{PhysRevB.71.035105}. 

This method was used previously to study Cr based materials such as CrX$_3$(X = Cl, Br, I) monolayers \cite{liu2016exfoliating}.
In this method, the linear behavior of the total energy with respect to the occupation number 
is imposed to correct the local and semi-local functionals. 
Prior to the implementation of the Linear Response method, the standard DFT calculation was first performed to obtain the converged charge. 
Following that, the interacting response of one single Cr atom was calculate by performing self-consistent DFT calculations with a series of 
Lagrange multipliers for the energy window from -0.08 to 0.08 eV, 
which usually falls within the linear region of number of d electrons versus Lagrange multipliers. 
The bare response of a single Cr atom was calculated by performing a non-self-consistent 
charge calculation with the same Lagrange multipliers as the self-consistent calculations. 
To avoid the interaction between the Cr atom and its periodic image within the unit cell, a 2$\times$2$\times$2 supercell was used during 
the U parameters calculations. 
The U parameter is then given by the difference between the second derivatives of the self-consistent energy, $\alpha^{scf}$ and the 
non-charge-self-consistent 
energy, $\alpha^{non-scf}$, with respect to the localized occupation of a single site\cite{PhysRevB.71.035105}, i.e.
\begin{equation}
	U = \frac{\partial \alpha_{i}^{scf}}{\partial q_{i}^{scf}} - \frac{\partial \alpha^{non-scf}_{i}}{\partial q_{i}^{non-scf}}
\end{equation} 
where $\alpha$ is the Lagrange multiplier, $q_i$ is the number of $d$ electrons of the single Cr atom. The first term in the right-hand side of 
the equation represents the interacting case, whereas the second term represents non-interacting case. 

\begin{figure}[htpb]
	\centering
	\includegraphics[width = 0.45\textwidth]{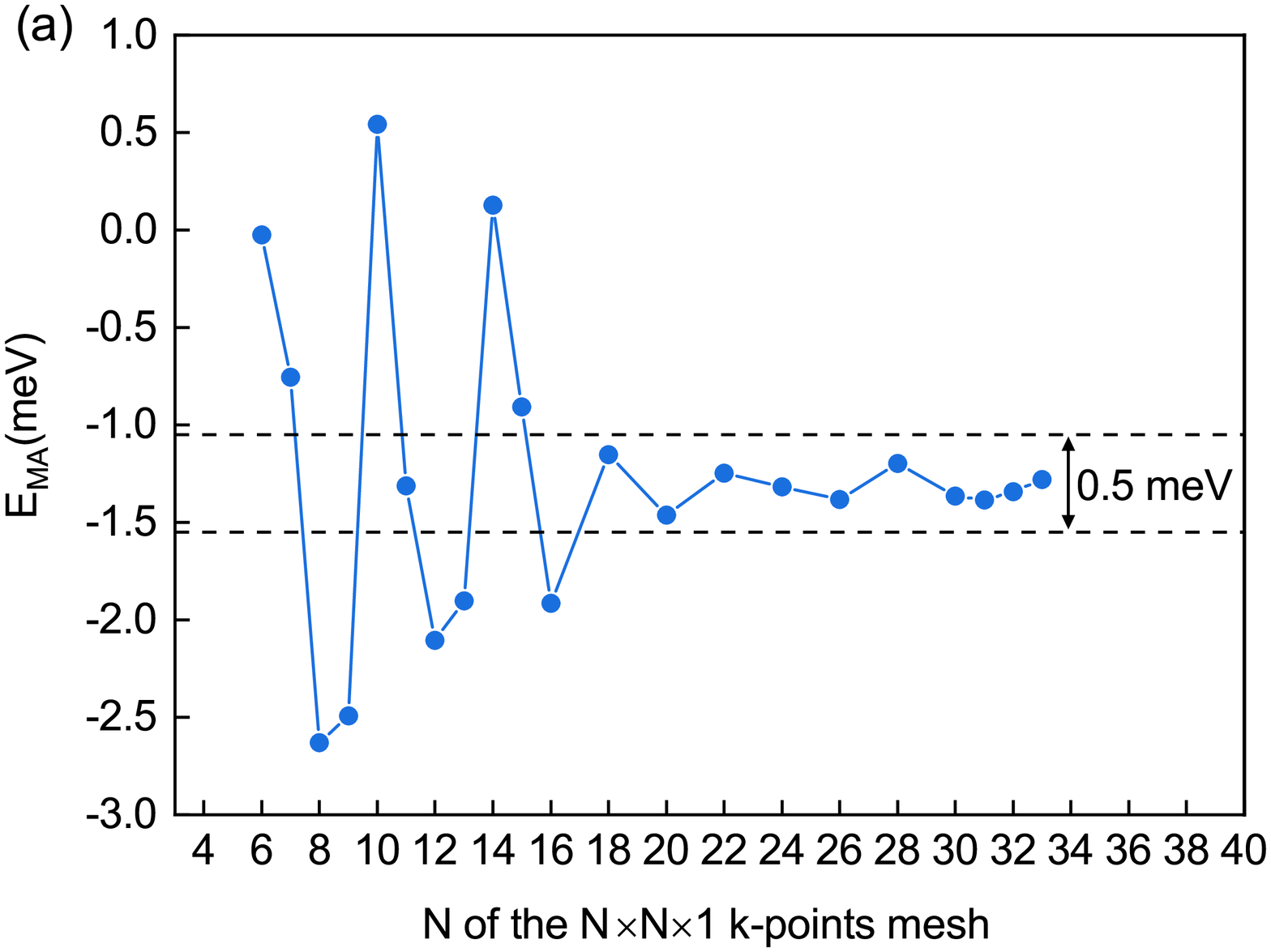}
	\includegraphics[width = 0.45\textwidth]{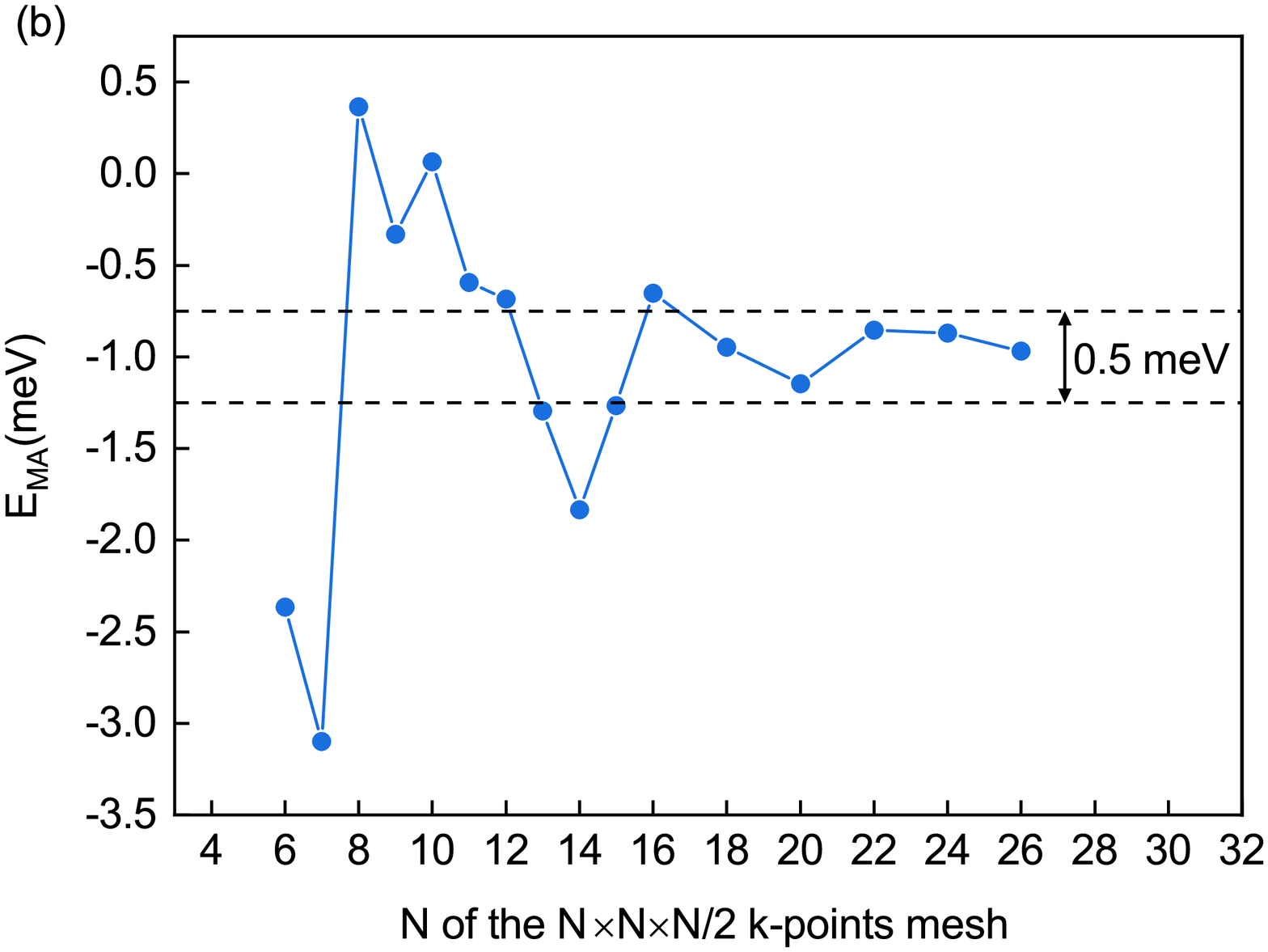}
	\caption{K-point convergence test on the magnetic anisotropy energy in (a) layered and (b) bulk 1T-CrTe$_2$}
	\label{fig:K_convergence}
\end{figure}

\begin{figure*}[htpb]
	\centering
	\includegraphics[width = 0.3\textwidth]{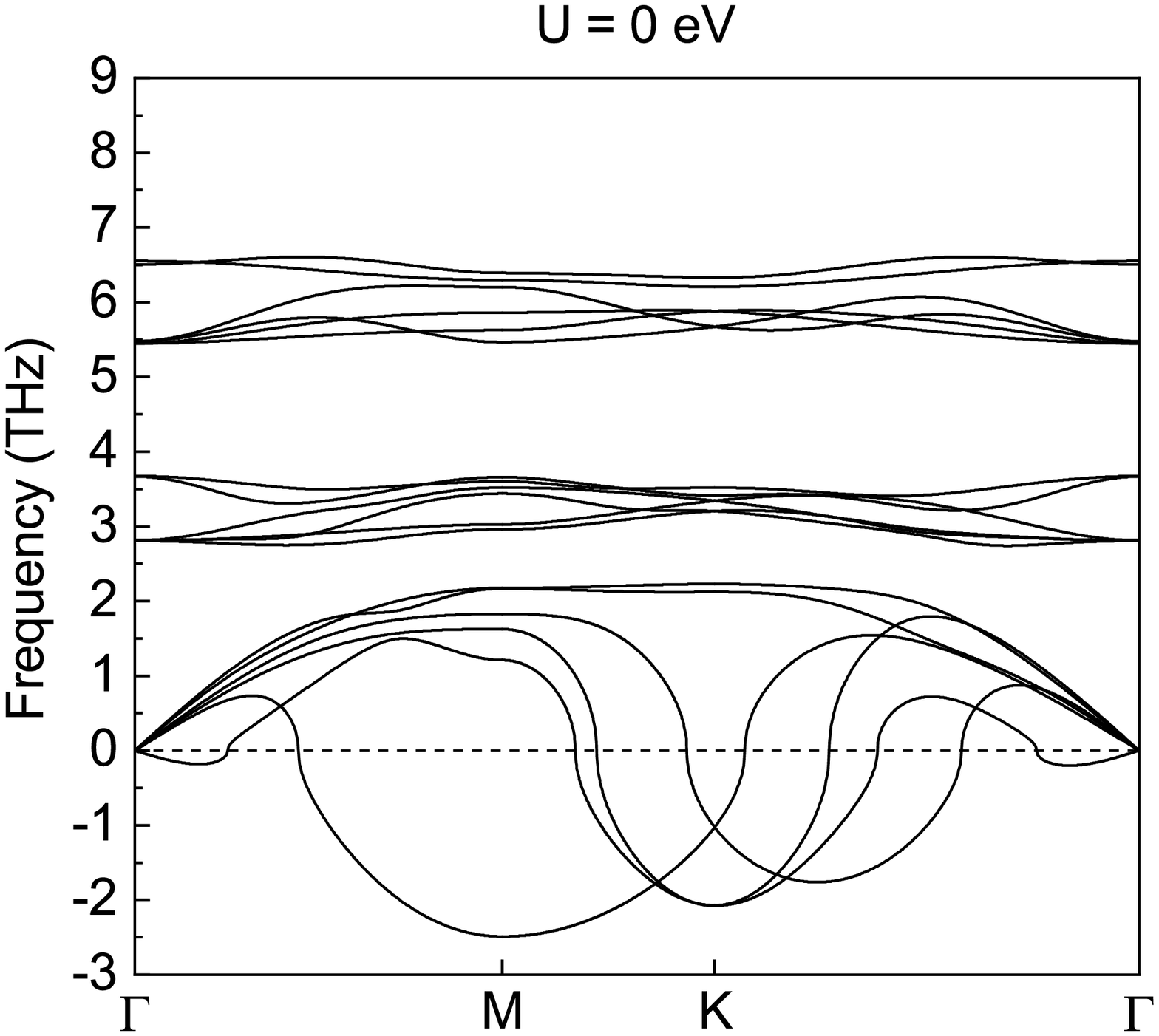}
	\includegraphics[width = 0.3\textwidth]{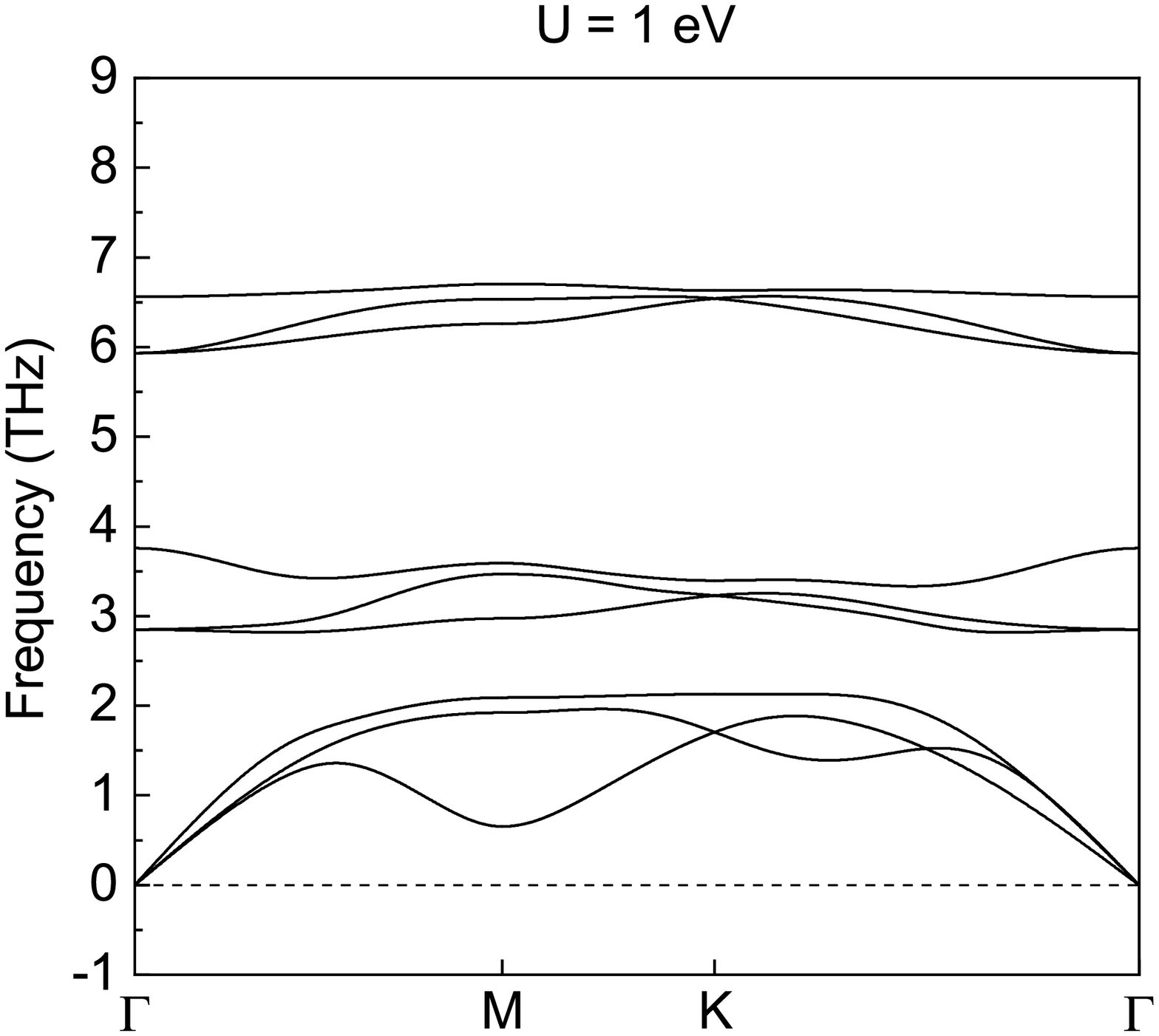}
	\includegraphics[width = 0.3\textwidth]{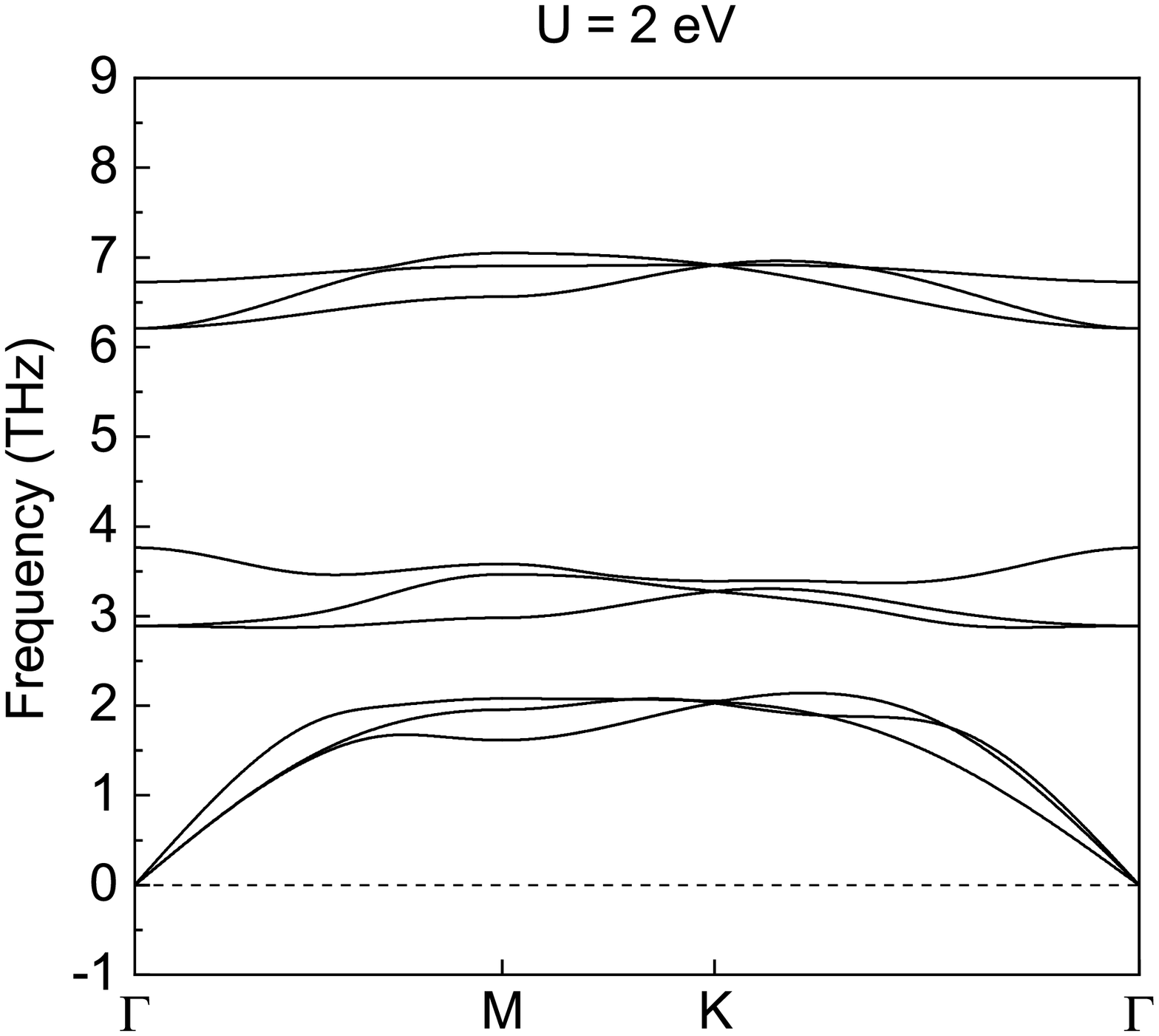}

	\includegraphics[width = 0.3\textwidth]{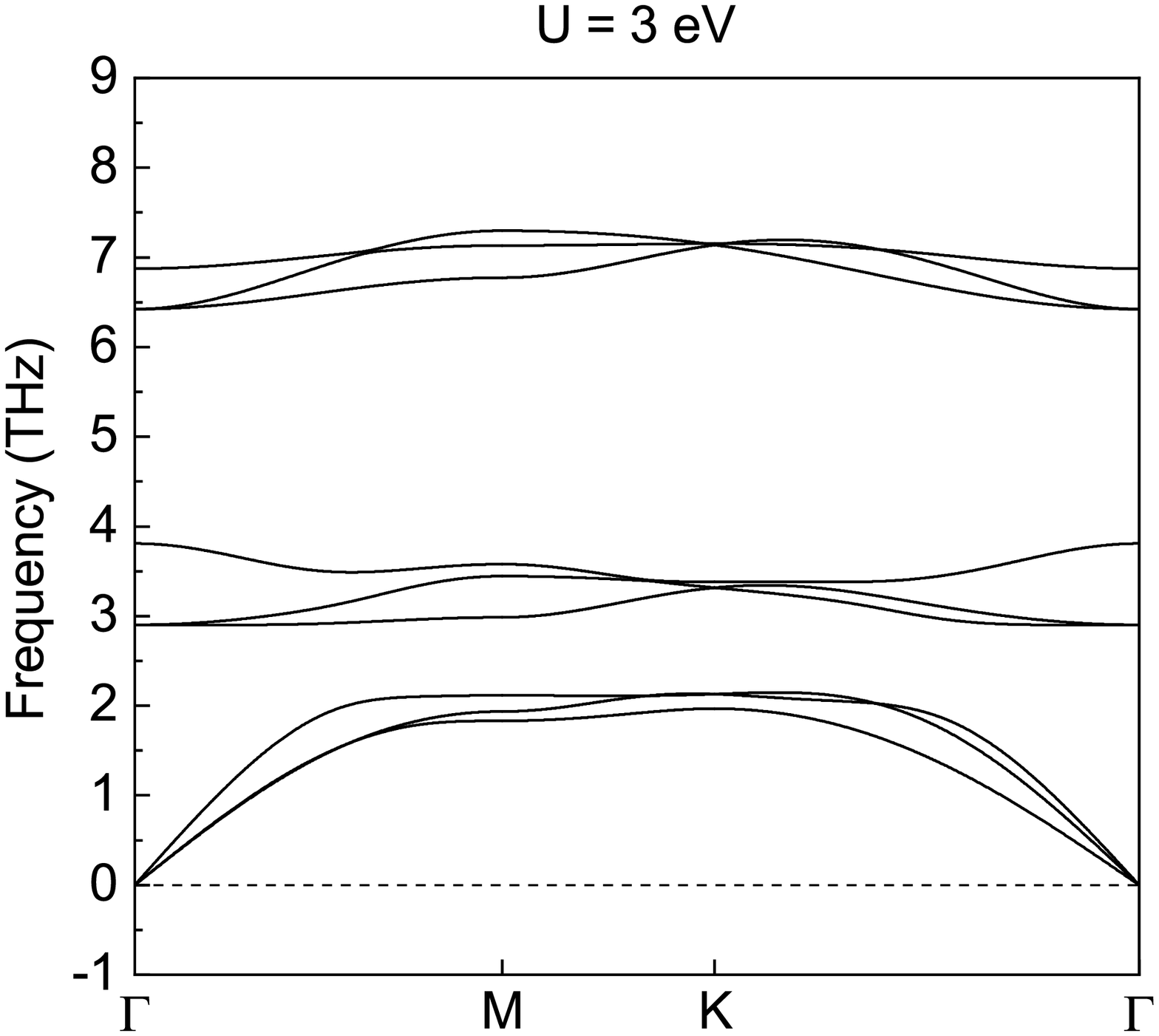}
	\includegraphics[width = 0.3\textwidth]{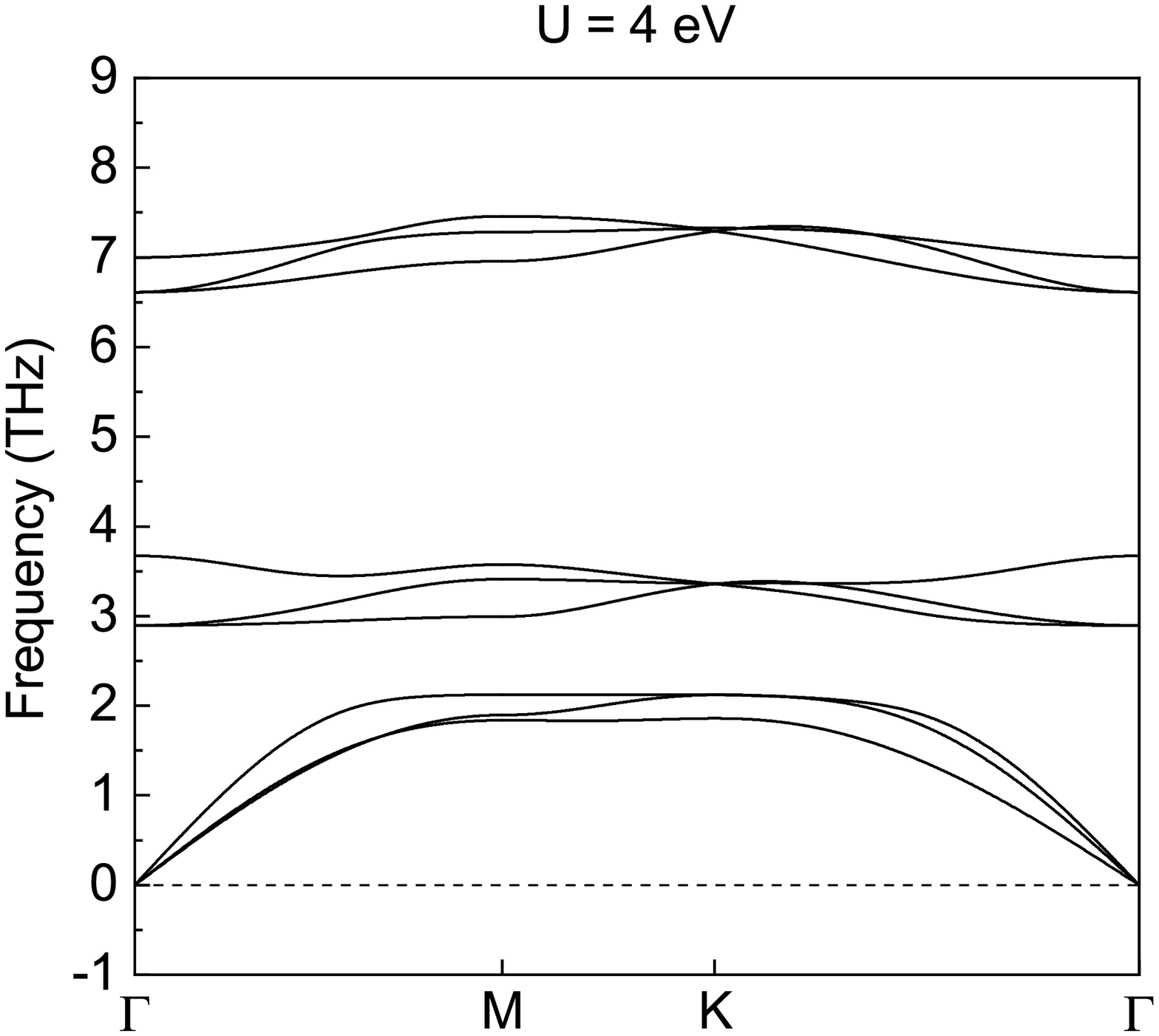}
	\includegraphics[width = 0.3\textwidth]{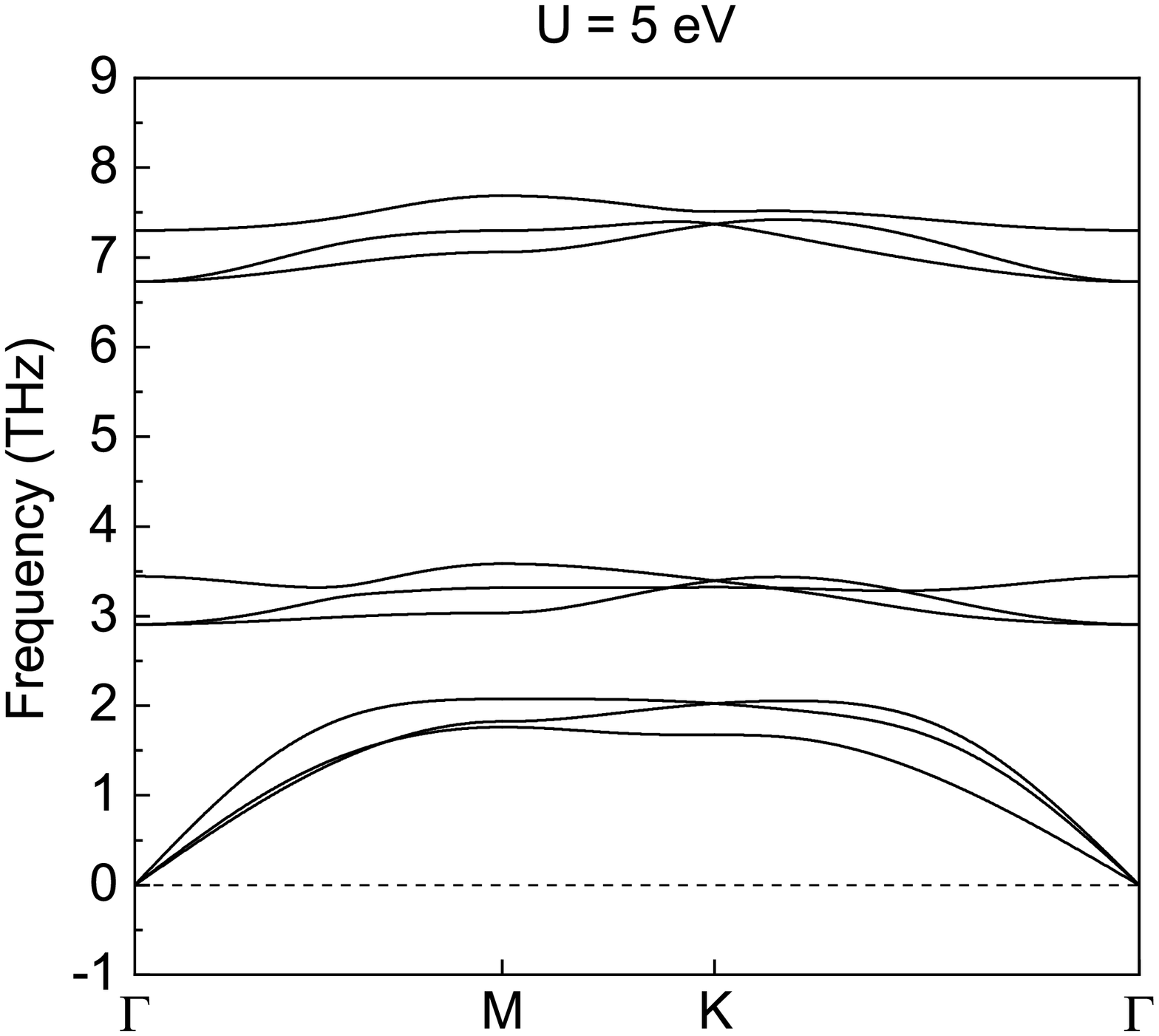}

	\includegraphics[width = 0.3\textwidth]{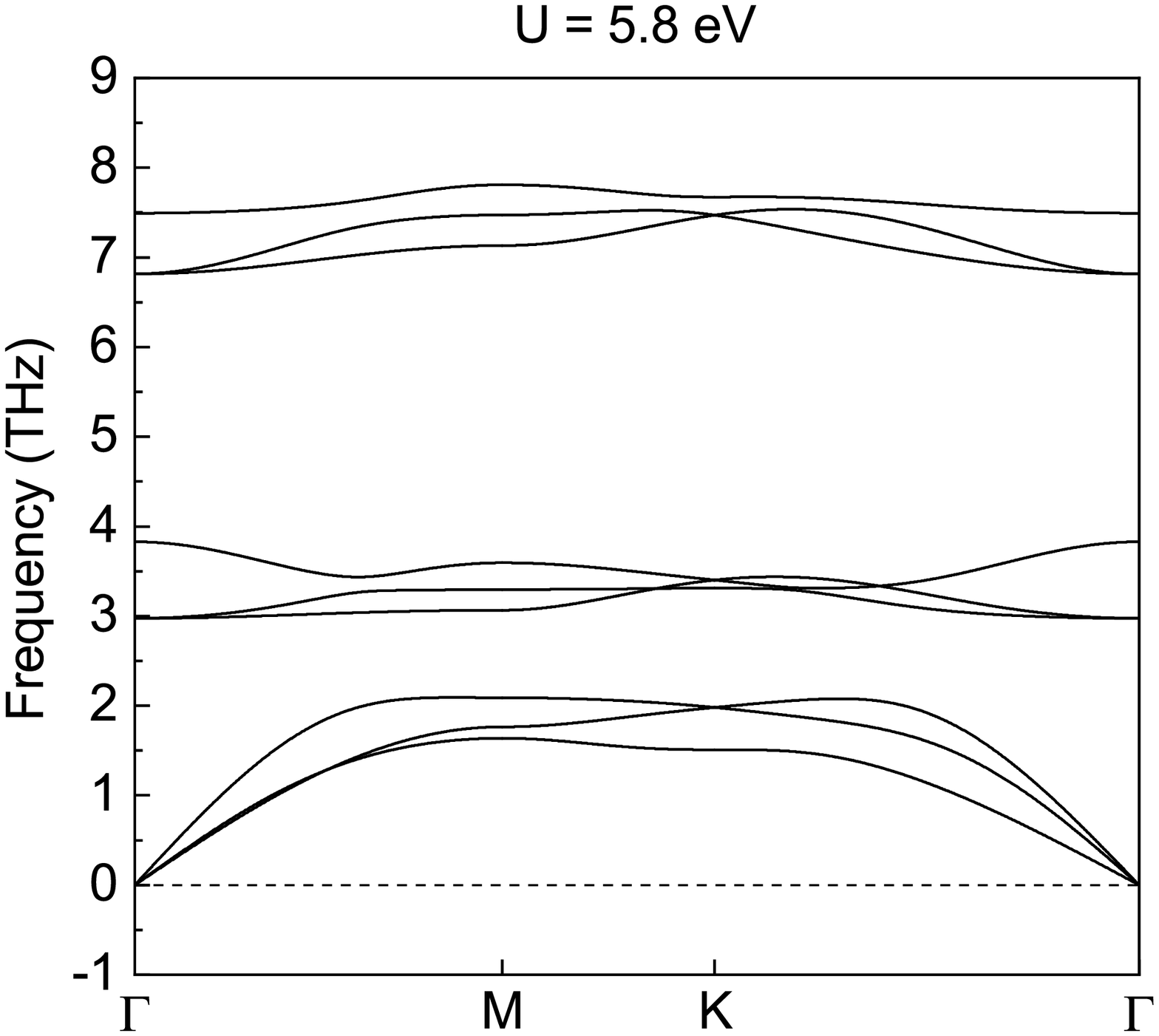}
	\includegraphics[width = 0.3\textwidth]{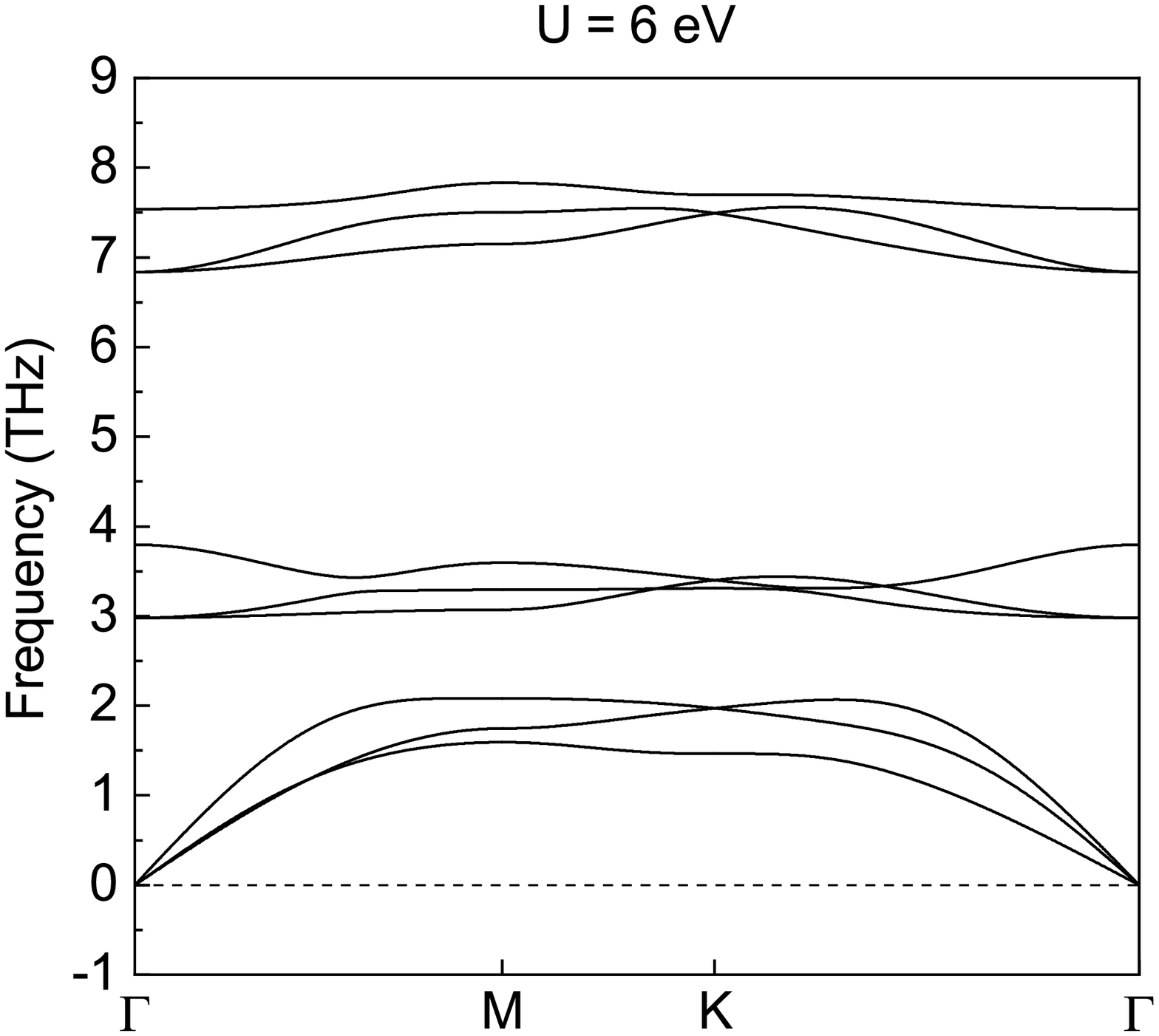}
	\includegraphics[width = 0.3\textwidth]{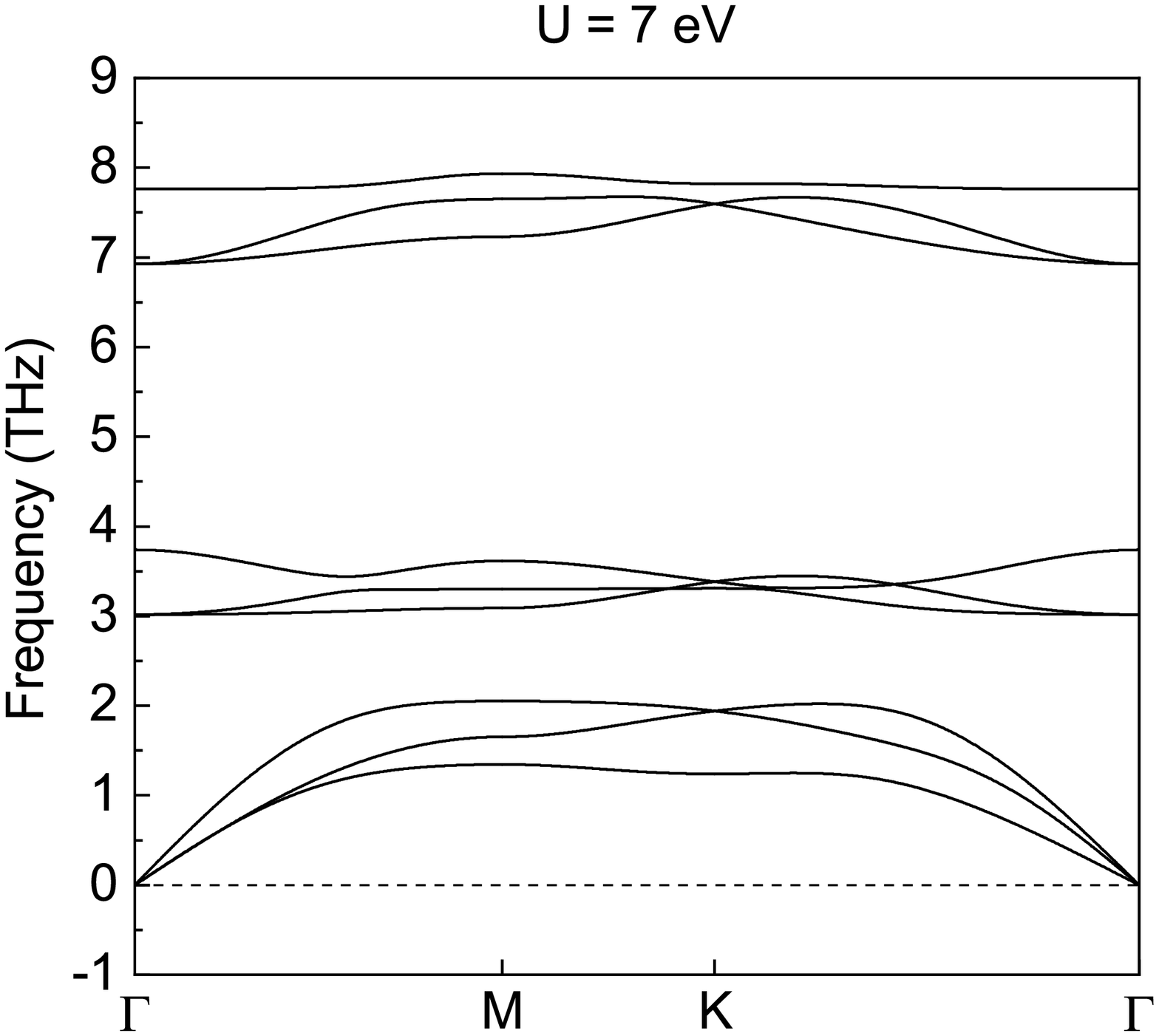}
	\caption{Phonon spectra of monolayer 1T-CrTe$_2$ calculated with different values of $U$.}
	\label{fig:Phonon}
\end{figure*}

\begin{figure}[htpb]
	\centering
	\includegraphics[width = 0.45\textwidth]{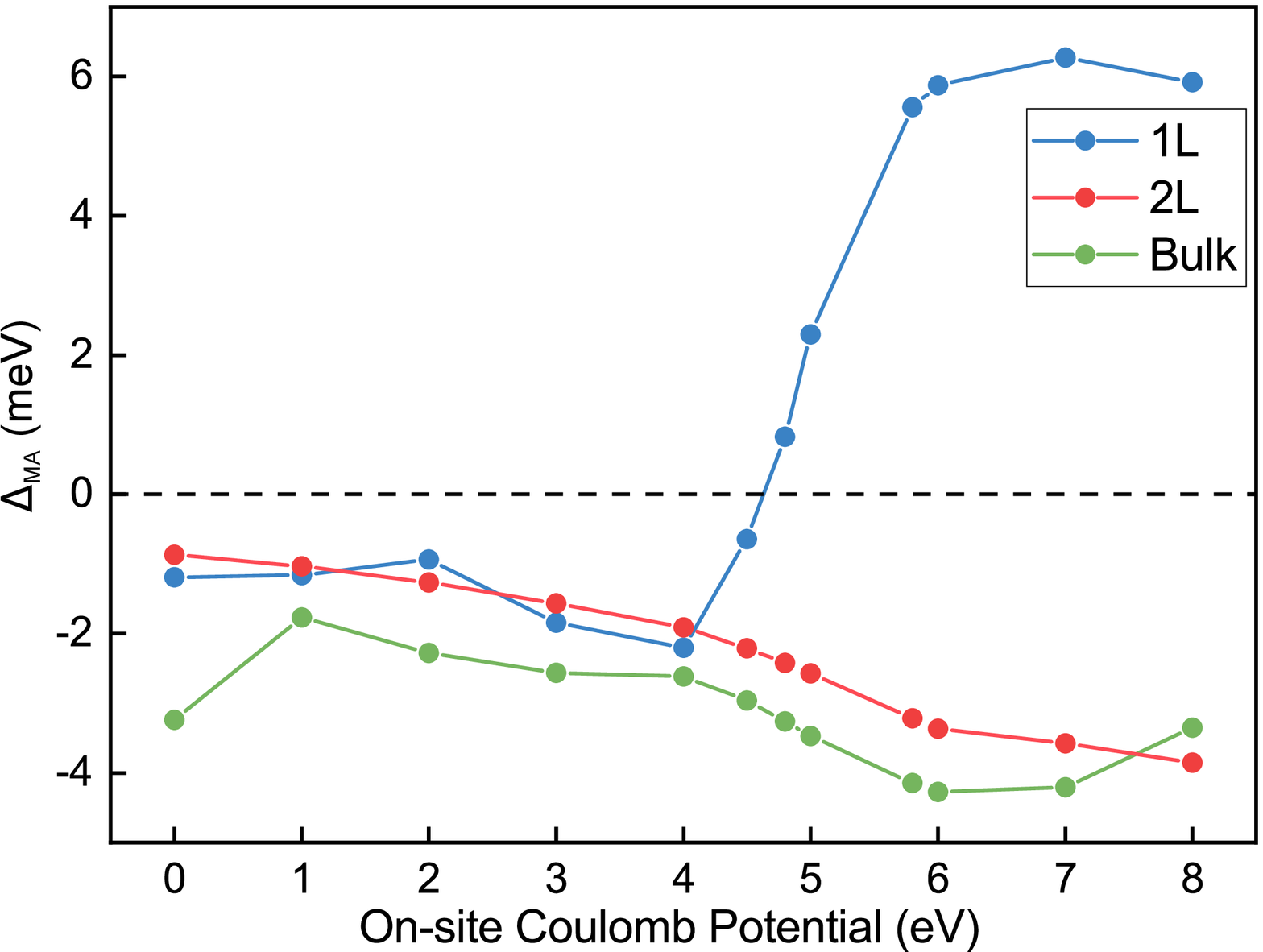}
	\caption{DFT calculated magnetic anisotropy energy ($\Delta_{\rm MA}$) as a function of on-site Coulomb potential $U$ for monolayer, bilayer, and bulk 1T-CrTe$_2$.}
	\label{fig:U_MAE}
\end{figure}

\noindent
{\bf Additional Data}\\
Additional data is provided below. 
Table \ref{tab:M} gives the magnetic moment per Cr atom for different values of $U$. 
Fig. \ref{fig:K_convergence} shows the k-point convergence of the MAE for monolayer and bulk \crte.
Fig. \ref{fig:Phonon} shows how the inclusion of $U$ stabilizes the phonon modes.
Fig. \ref{fig:U_MAE} shows the dependence of the MAE on the value of $U$.

\begin{table*}[htbp]
	\renewcommand\arraystretch{1.25}
	\caption{DFT calculated magnetic moment ($\mu$B) per Cr atom in 1T-CrTe$_2$ with different $U$ values.}
	\begin{ruledtabular}
		\begin{tabular}{|l|l|l|l|l|l|l|l|l|l|l|l|l|}
			\hline
			U (eV)     & 0     & 1     & 2     & 3     & 4     & 4.5   & 4.8 & 5     & 5.8   & 6     & 7     & 8     \\ \hline
			Mono-layer & 2.390 & 2.548 & 2.667 & 2.769 & 2.860 & 2.913 & 2.953 & 2.978 & 3.085 & 3.110 & 3.208 & 3.285 \\ 
			Bi-layer   & 2.406 & 2.562 & 2.692 & 2.806 & 2.903 & 2.949 &  2.976   & 2.993 & 3.059 & 3.074 & 3.148 & 3.216 \\
			Bulk       & 2.385 & 2.528 & 2.661 & 2.781 & 2.884 & 2.933 &  2.958  & 2.981 & 3.053 & 3.070 & 3.147 & 3.219 \\
		\end{tabular}
	\end{ruledtabular}
	\label{tab:M}
\end{table*}

\clearpage
\newpage

\noindent
{\bf Derivation of magnon dispersion}\\

Performing the Holstein-Primakoff transformation \cite{Holstein_Primakoff1940}
defined by the operator substitution
$S - \hat{n}_j = S_j^z$ where $\hat{n}_j = a_j^\dagger a_j$ is the magnon number operator, 
and keeping terms up to first order in $\hat{n}_j$, Eq. (4) of the main text becomes,

\begin{align}
H &= \left( -J_z S^2 Z - K_u S^2 + g \mu_B B_z S \right) 
\nonumber \\
& - 
\tfrac{S}{N} \sum_j \sum_\delta J_{xy} \left( a_j^\dagger a_{j+\delta} + a_{j+\delta}^\dagger a_j \right) 
\nonumber \\
& + \tfrac{2S}{N} \sum_j 
\left( J_z Z + K_u - \tfrac{g \mu_B B_z}{2S}\right) 
a_j^\dagger a_j 
\label{eq:AH_HP_j}
\end{align}

where $Z = 6$ is the number of nearest neighbor Cr atoms, and $\delta$ represents the 
position of the 6 nearest neighbors with respect to site $j$.
The first term is the energy with all spins aligned along $z$,
i.e. $S_j^z = S = 3/2$, and the next two terms are the change in energy
due to the change in $S^z$ resulting from the presence of magnons.
Substituting in the Fourier representations of the operators, 
$a_j = \frac{1}{\sqrt{N}} \sum_\kv e^{-i\kv \cdot \Rv_j} a_\kv$, 
into the second and third terms
gives the Hamiltonian governing the magnon dispersion,
\begin{equation}
H_{m} = \tfrac{2S}{N} \sum_{\kv} \left[ J_z Z + K_u - \tfrac{g\mu_B B_z}{2S} - J_{xy} {\rm Re}\{f(\kv)\} \right] a^\dagger_\kv a_\kv
\label{eq:H_AH_magnon_2}
\end{equation}
where 
$f(\kv) \equiv \sum_\delta e^{-i\kv \cdot \delta} \in \mathbb{R}$ 
is the form factor resulting from the
sum over the 6 nearest Cr neighbors.
\\

\noindent
{\bf Relation between the experimental value for $\Kexp$ and the parameter $\Keff$ used in the
Heisenberg Hamiltonian}\\

The expression $\Kexp = -\frac{\mu_0 H_s M_s}{2}$ used to extract $\Kexp$ from experimental magnetization 
curves \cite{zhang2021room}
is derived from the formula for the angular dependence of the energy of
a magnetic thin film in the presence of an applied field $H$,
$E = -\Kexp \cos^2 (\theta) - \mu_0 M_s H \cos(\phi - \theta)$, where $\phi$ and $\theta$ are the polar angles of
the magnetic field and the magnetization, respectively \cite{1996_MJohnson_MAE_RepProgPhys}. 
Minimizing $E$ with respect to $\theta$, and
letting $H$ be in-plane, so that $\phi = \pi/2$, gives $\Kexp = -\frac{\mu_0 H_s M_s}{2}$,
where $H_s$ is the saturation field.
The angular dependence of our Heisenberg Hamiltonian with an arbitrary direction for $B$ comes from the two terms,
$-\Keff (S^z)^2 + g \mu_B \mu_0 \Hv \cdot \Sv$. 
Writing $S$ as a classical vector gives $-\Keff S^2 \cos^2(\theta) - \mu_0 H M_s \cos(\phi-\theta)$ 
where $\Mv_s = -g \mu_B \Sv$.
Minimizing as before gives $\Keff = -\frac{\mu_0 H_s M_s}{2S^2}$, so that $\Keff = K_{\rm exp} / S^2$.
\\

\providecommand{\noopsort}[1]{}\providecommand{\singleletter}[1]{#1}%
\end{document}